\documentclass[sigconf]{acmart}
\pdfoutput=1

\AtBeginDocument{%
  }

\setcopyright{acmlicensed}
\copyrightyear{2025}
\acmYear{2025}
\acmConference['WWW']{Make sure to enter the correct
  conference title from your rights confirmation emai}{April 28--May 2,
  2025}{Sydney, Australia}
\usepackage{graphicx}
\usepackage{caption}
\usepackage{epstopdf}
\usepackage{subfigure}
\usepackage{nicefrac}

\DeclareMathAlphabet{\mathbcal}{OMS}{cmsy}{b}{n}
\usepackage{mathrsfs}
\usepackage{comment}

\usepackage{bm}
\usepackage{bbm}
\usepackage{paralist}
\usepackage{graphicx}
\usepackage{csvsimple}
\usepackage{booktabs}
\usepackage[font=small]{caption} 

\usepackage[strict]{changepage}

\usepackage{xcolor}

\usepackage{placeins}

\definecolor{formalshade}{rgb}{0.95,0.95,1}
\definecolor{darkblue}{rgb}{0.0, 0.0, 0.55} % Define darkblue color

\usepackage{framed}

\definecolor{formalshade}{rgb}{0.95,0.95,1}

\newenvironment{formal}{%
  \MakeFramed{\advance\hsize-\width\FrameRestore}%
  \noindent\hspace{-4.55pt}% disable indenting first paragraph
  \begin{adjustwidth}{4pt}{7pt}%
}
{%
  \end{adjustwidth}\endMakeFramed%
}

\usepackage{colortbl}

\begin{document}

% \title{LLMs are good predictors of ...}
\title[Luera et al.]{Optimizing Data Delivery: Insights from User Preferences on Visuals, Tables, and Text}

% Reuben Luera, Ryan Rossi, Franck Dernoncourt, Alexa Siu, Sungchul Kim, Tong Yu, Ruiyi Zhang, Xiang Chen, Nedim Lipka, Zhehao Zhang, Seon Gyeom Kim, Tak Yeon Lee 

\author{Reuben Luera}
\affiliation{%
  \institution{University of California--San Diego}
  \city{San Diego}
  \state{California}
  \country{USA}}
  \email{raluera@ucsd.edu}

\author{Ryan Rossi}
\affiliation{%
  \institution{Adobe Research}
  \city{San Jose}
  \state{California}
  \country{USA}}
  \email{ryrossi@adobe.com}

\author{Franck Dernoncourt}
\affiliation{%
  \institution{Adobe Research}
  \city{Seattle}
  \state{Washington}
  \country{USA}}
\email{dernonco@adobe.com}

\author{Alexa Siu}
\affiliation{%
  \institution{Adobe Research}
  \city{San Jose}
  \state{California}
  \country{USA}}
  \email{asiu@adobe.com}

\author{Sungchul Kim}
\affiliation{%
  \institution{Adobe Research}
  \city{San Jose}
  \state{California}
  \country{USA}}
  \email{sukim@adobe.com}

\author{Tong Yu}
\affiliation{%
  \institution{Adobe Research}
  \city{San Jose}
  \state{California}
  \country{USA}}
  \email{tyu@adobe.com}

\author{Ruiyi Zhang}
\affiliation{%
  \institution{Adobe Research}
  \city{San Jose}
  \state{California}
  \country{USA}}
  \email{ruizhang@adobe.com}

\author{Xiang Chen}
\affiliation{%
  \institution{Adobe Research}
  \city{San Jose}
  \state{California}
  \country{USA}}
  \email{xiangche@adobe.com}

\author{Nedim Lipka}
\affiliation{%
  \institution{Adobe Research}
  \city{San Jose}
  \state{California}
  \country{USA}}
  \email{lipka@adobe.com}

\author{Zhehao Zhang}
\affiliation{%
  \institution{Dartmouth College}
  \city{Hanover}
  \state{New Hampshire}
  \country{USA}}
  \email{zhehao.zhang.gr@dartmouth.edu}

\author{Seon Gyeom Kim}
\affiliation{%
  \institution{KAIST}
  \city{Daejeon}
  \country{South Korea}}
  \email{ksg_0320@kaist.ac.kr}

\author{Tak Yeon Lee }
\affiliation{%
  \institution{KAIST}
  \city{Daejeon}
  \country{South Korea}}
  \email{takyeonlee@kaist.ac.kr}
%Daejeon

\begin{abstract}
In this work, we research user preferences to see a chart, table, or text given a question asked by the user. This enables us to understand when it is best to show a chart, table, or text to the user for the specific question. For this, we conduct a user study where users are shown a question and asked what they would prefer to see and used the data to establish that a user's personal traits does influence the data outputs that they prefer. Understanding how user characteristics impact a user's preferences is critical to creating data tools with a better user experience. Additionally, we investigate to what degree an LLM can be used to replicate a user's preference with and without user preference data. Overall, these findings have significant implications pertaining to the development of data tools and the replication of human preferences using LLMs. Furthermore, this work demonstrates the potential use of LLMs to replicate user preference data which has major implications for future user modeling and personalization research.
\end{abstract}




\received{20 February 2025}
\received[revised]{12 March 2025}
\received[accepted]{5 June 2025}

\maketitle
\fancyhead[L]{Optimizing Data Delivery} % Left side of the header
\fancyhead[R]{Luera et al.} % Right side of the header

\section{Introduction}

As data and large language models (LLMs) continue to grow in prominence, it is crucial to identify the most effective ways to present data outputs, as the format, whether chart, table, or text, significantly influences how users engage with and interpret information \citep{tufte1983visual, few2004show}. With datasets becoming larger and more complex, visualizations are increasingly necessary to help users digest the information effectively \citep{godfrey2016interactive}. The expansion of LLM use in data analysis adds another layer, making it essential to understand when these models should present different output formats. Moreover, individuals have varying preferences for data representation, driven by their unique characteristics, such as experience with data analysis and visualization, age, and work experience. This paper investigates these preferences, exploring how user characteristics shape their choice of data outputs, and how LLMs can adapt to deliver more personalized and intuitive results \citep{DBLP:journals/corr/abs-2005-14165}. Ultimately, by dynamically tailoring outputs based on user backgrounds, LLMs can offer a more customized and effective experience, helping users better understand and utilize data.

\noindent\textbf{Summary of Main Contributions.} 
The key contributions of this work are as follows:

\begin{compactitem}
\setlength{\itemindent}{0pt}
    \item \textbf{A comprehensive user survey and methods design.} We outline the key components of the Amazon Mturk survey, detailing the respondent population, survey setup, specific user and data-related questions, and the instructions provided to participants.
    \item \textbf{An analysis of general data output preference results.} The first research question examined the general population’s preferred data output for a given question, aiming to establish a baseline for data preferences without considering user characteristics.
    \item \textbf{An overview of data output preferences when organized by personal user characteristics.} The 2nd research question explored how a user’s personal characteristics influence their data output preferences, focusing on experience with data analysis and visualization and their age.
    \item \textbf{An overview of data output preferences when organized by work experiences.} The third research question explored how work experience, including industry and role, influences users’ data output preferences.
    \item \textbf{A comparison between human and GPT preferences.} We used GPT to see if it could predict the human preference data we received throughout the study
    
\end{compactitem}

\section{Related Work} \label{sec:related-work}

\subsection{Text to Visualization Generation}

Visualization generation, whether charts or tables, from natural language, has become increasingly common as LLMs and natural language interfaces (NLIs) for data grow in popularity. These systems allow users with less data literacy to create comprehensive charts and expansive tables. Current research in this area often focuses on the creation of these systems \citep{tian2024chartgpt, rashid2021text2chart, narechania2020nl4dv}, but does not do expansive research on which data output is best given a natural language question or a user's individual characteristics. 

In ChartGPT, \cite{tian2024chartgpt} explains their system as an LLM that is capable of generating charts given abstract natural language inputs. ChartGPT is effective at grouping parts of the natural language inputs into subtasks to identify key parts of it to present an appropriate visualization. Similarly, \cite{rashid2021text2chart}'s Text2Chart uses BERT and LSTM, two deep learning models, to also create visualizations from natural language. Meanwhile, \cite{narechania2020nl4dv}'s NL4DV relies on traditional NLP methods, like dependency analysis and lexical parsing, instead of on an LLM. We aim to take this research further by simulating the natural language interactions in these systems and exploring data output mediums while also considering different user characteristics. 
%\subsection{Figure Captioning \& Text Summarization of Visualizations}

\vspace{-1.8mm}
\subsection{Visualization + Text for Analysis}
A lot of work has been done to test the varying degrees in text and data visualizations can be used in tandem with one another to help users digest data. Systems like Eviza \cite{setlur2016eviza} create visualization and text combinations that make it easier for users to understand the data they are dealing with. Meanwhile, systems like \cite{smitsaltgosling, singh2024figura11y} help users create alt-text for a given data visualization. Creating text for a given data visualization helps users understand data visualizations that may have been inaccessible to them for whatever reason. On the other hand, \cite{hearst2019would} found that 40\% of users do not prefer to see charts when in a conversational UI. Instead, they prefer their answers to be outputted in text.

Work by~\cite{stokes2022striking} investigated the value of including varying degrees of textual information for understanding univariate line charts.
This work focused only on univariate line charts and also investigated the placement of text and its impact.
While that work investigated showing only a line chart, a line chart with visual and short text annotations, along with showing the user a longer text description.
It did not investigate the combination of showing the user a line chart with a detailed text description, nor do they study the utility of including a table with raw data.
Furthermore, the findings of that work also conflated visual annotations, as the intermediate charts included in their study, include both visual annotations (i.e. highlighting the maximum value of a time series, and then displaying a textual annotation near it) as well as short textual annotations that are placed near the highlighted point on the line chart.

\vspace{-1.8mm}

\subsection{Accessibility \& Technical Literacy}

The capability to display data in several different formats, such as in charts, tables, or text, is significant for accessibility reasons. As data becomes more relevant to different sectors, data illiteracy can be a limiting factor \citep{disseldorp2020data,vemulapalli2024overcoming, d2017creative}. Moreover, taking physical and neurological disabilities into account ensures that these systems are more accommodating \citep{lundgard2021accessible,lee2024inclusive, wu2023empowering}. Overall, by researching how a person's unique characteristics impact their data preferences, data visualization applications can become more accessible.

Taking data literacy issues a step further, many users in non-technical fields often find themselves having to interact with complex data sets and visualizations \citep{vemulapalli2024overcoming}. Furthermore, many companies have limited resources to teach their technically limited workers how to use data appropriately \citep{disseldorp2020data}, and often have to rely on common end-of-year performance reviews to gauge a worker's technical literacy. For most companies and employees, by this point, it is often too late. Again, most companies do not have the resources to bring their employees up to speed on data techniques, so research continues to be needed for an alternative. Given this, conducting research that takes varying technical literacy and disabilities into account can help understand how to best serve those who are often marginalized in data conversations.

\begin{figure*}
    \centering
    \vspace{-2mm}
    \includegraphics[width=.77\linewidth]{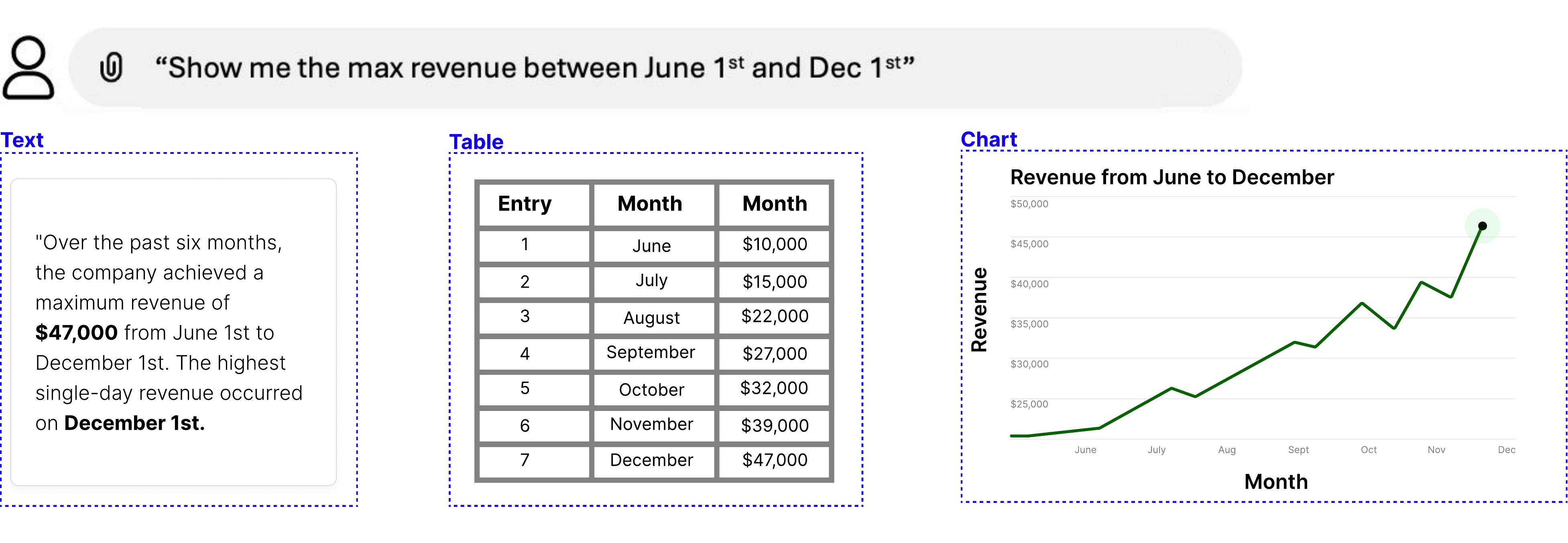}
    \vspace{-4mm}
    \caption{
    When answering the user survey, MTurkers were shown this figure as an example of what the data outputs could potentially look like. The leftmost example is what the text answer would look like, the middle is the answer in the form of a table, and the right is the answer in the form of a chart. Then they were asked, "Given a data analysis question, is it most useful to show the user text, data table, or chart?"}
    \label{fig:Instructions}
    \vspace{-2mm}
\end{figure*}

\section{Study} \label{sec:study}
We aim to conduct a user study, focused on when it is best to show a visualization, table, text, or any combination of these options to the user for a given question.
Our study consists of a user survey and a pre-survey questionnaire. The survey will ask a question and then prompt the user to choose between a chart, text, or table result. We will conduct this user study collecting user preference data and synthesizing these results, highlighting interesting trends.  

In this work, we study the following research questions:

\noindent
\textbf{RQ1: }Given a data question, in general, will users prefer to see the answer visualized as a table, text, or chart? 

\begin{itemize}
    \item \textbf{H1} users will prefer to see their data represented as charts, tables, and then text in that order.
\end{itemize}

\noindent
\textbf{RQ2:} Are there certain personal user characteristics that correlate with the users' preference to see a chart, table or text?

\begin{itemize}

    \item \textbf{H2a}: Respondents with more data visualization experience will prefer charts, while users with less experience will show a stronger preference towards tables and text.

    \item \textbf{H2b}: Respondents with more data analysis experience will also prefer charts over text and table outputs.

    \item \textbf{H2c}: In terms of age, younger respondents will show a stronger preference for charts while older respondents will prefer tables and text.
  
\end{itemize}

\noindent
\textbf{RQ3:} Does a respondent's role at work or industry they work in correlate with their preference to see a visualization, table, or text?
\begin{itemize}

    \item \textbf{H3a}: Respondents will prefer different data outputs based on the role they play at work, with more presentation-oriented roles preferring charts.

    \item \textbf{H3b}: Respondents will prefer different data outputs based on the industry they work in, with more technical industries preferring tables and text.
\end{itemize}

\noindent
\textbf{RQ4:} Can LLMs be used to predict whether a question should be answered with a visualization, data table, or text?
\begin{itemize}
    
    \item \textbf{H4a}: LLM Alignment with Humans: Using only an LLM without any user-specific personalization will perform poorly on predicting whether a user should be answered with a visualization, data table, or text.
   
    \item \textbf{H4b}: Personalized LLMs: Does including user-specific examples and user characteristics in the LLM improve the accuracy of the LLM in generating the preferred answer for individual users?
   
    \item \textbf{H4c}: User-specific Accuracy: Can the personalized LLM approach perform well for some users and worse for others?
\end{itemize}

\subsection{Participants}

In order to conduct the survey, we used Amazon Mechanical Turk. In total we uploaded 5 different sets of questions, each set having approximately 50 questions each in addition to the eight demographic questions at the beginning. The data questions were created in a different survey on Upwork, where we asked participants to create general data questions that a user might ask an LLM. In doing so, we pulled from that bank of questions and select them based on their relevance to our survey. Given that there were 5 sets of questions, and 50 questions per set, we essentially had 250 unique questions in total. For each set of questions, the users were compensated \$1.40.

For every set, we initially recruited 200 respondents per set. At fifty questions per set, we essentially had about 50,000 unique responses. From there, we cleaned out any responses were subpar or were suspected as duplicate responses. We also set a certain time threshold (400 seconds) and cleaned out any user that fell beneath this threshold as our survey could not reasonably be completed in less than 400 seconds.
 
\vspace{-1mm}
\subsection{Method}
The survey was broken down into three subsections: the demographic questions, the instructions, and the data survey questions. The answers from the two sections were used in tandem to identify trends and correlations. The same instructions were presented to each user, ensuring consistency with every survey. 

\vspace{-1mm}
\subsubsection{User Characteristic Questions}
In order to answer RQ1 and gain a better understanding of if and how a person's characteristics impact their data output preference, we first had to gather user characteristics from each respondent. In doing so, we can map which, if any, user characteristics impact a person's data output and which do not. The questions asked at the beginning of the survey can be found in section \ref{demoQs}.

\vspace{-1mm}
\subsubsection{Instructions} 

After the user characteristic questions, we showed a consistent figure that presented an example scenario (Fig. \ref{fig:Instructions}). As seen in Fig. \ref{fig:Instructions} the figure consists of a given prompt and then shows three examples of the potential data output mediums. As the user goes through the rest of the survey, they can reference this figure for examples of the different output mediums. 

Furthermore, it was decided that the instructions would be the only place where the user could see examples of text, tables, and charts. This was intentionally done so as not to bias the respondent on each question. If the respondent saw a specific text, table, or chart for each question, this could potentially disrupt their own personal beliefs on what the output could look like. While we understand the drawbacks of this approach, this was the best way to mitigate bias. Given these reasons, we maintained that the instructions would be succinct and section that respondents can reference to get examples of what type of data output they would prefer.

\vspace{-1mm}
\subsubsection{Survey Questions}

As mentioned, each survey had 50 unique questions that all had the same question structure. Each question begins with the same basic instruction: "Given the question below, please select your preference on how the answer to the question should be presented." After this instruction, the user is presented with a generic prompt and asked to choose what data output medium would best fit the needs of that given question (Fig \ref{fig:Instructions}). 

After being shown these questions, users were presented with three options: text, table, or chart. Given the question, the users were tasked with choosing which of these three data output methods they preferred. These options were presented as radial buttons and the respondents were tasked with choosing one of the three options for each of the 50 questions.

\begin{figure}[t]
\centering
\includegraphics[width=.9\linewidth]{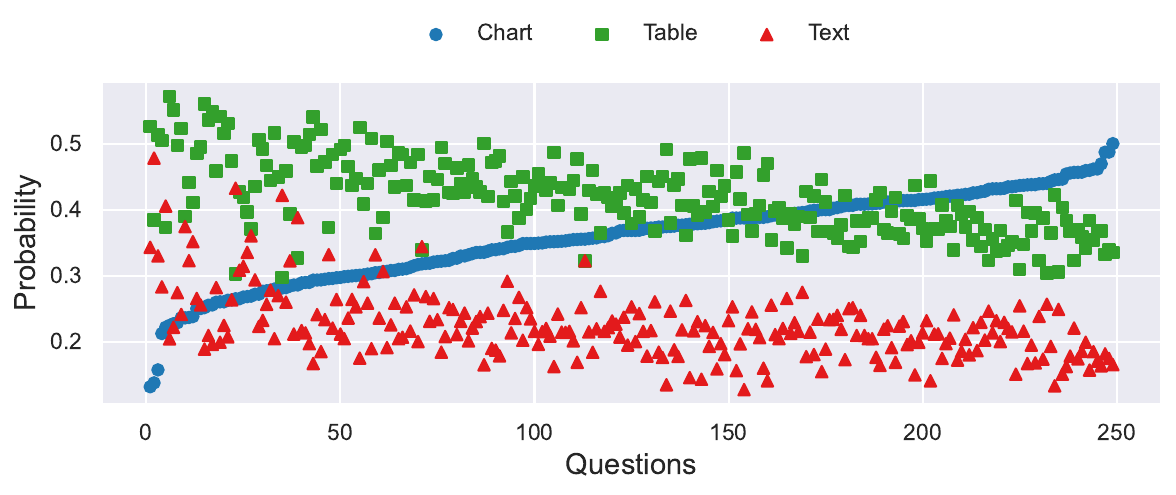}
\vspace{-2mm}
\caption{%
% \textbf{Question Responses}
For each question, we aggregate all the user preferences, and derive a distribution, which is shown above (sorted by the value for chart, which is why we see a nice curve for the probability of chart).
Notably, we see that as the probability that a user prefers a chart increases, the probability a user prefers text or table decreases.
% We also observe that in all cases, most users 
} \label{fig:question-distribution}
\vspace{-2mm}
\end{figure}

\section{Results} \label{sec:results}

\subsection{RQ1: General Preferences}

\medskip\noindent\textbf{Findings:} RQ1 asks: Given a data question or prompt will users prefer to see the answer visualized as a table, text, or chart?
From the results, we found that the most common preference was for tables at 41.70\%, with charts at 36.2\%, and text preferred far less at 21.97\% (Fig. \ref{fig:RQ1Chart}).

Given this, it was clear that there was a large amount of variability between the three data output preferences. Figures \ref{fig:question-distribution} \& \ref{fig:user-distribution} illustrate how user preferences are distributed across individual users and questions. Looking at Figure \ref{fig:question-distribution}, it is clear that users exhibited preferences for all three data output types to some degree. This signifies the variability within the preferences and speaks to the complexity of the task. The user preferences did not have a uniform consistency and instead displayed significant variation. On a similar note, Figure \ref{fig:user-distribution} shows that there was a wide distribution given individual questions, following a similar pattern to Figure \ref{fig:question-distribution}. Overall, this variability in preferences underscores the need for a deeper analysis on ways to personalize outputs based on user characteristics.

\medskip\noindent\textbf{Analysis Details \& Discussion:}

In our original hypothesis (H1) it was stated that \textit{H1: Users will prefer to see their data represented as charts, tables, and then text in that order.} 

This hypothesis was \textbf{partially correct} as the text was the least preferred output format, but unlike our hypothesis tables and charts were the most preferred outputs, respectively. These results were gathered by calculating the percentage of each output answer within the larger dataset.

\begin{enumerate}
    \item \textbf{Tables} were the most preferred data output, with 41.7\% of responses preferring tables. This preference may have been caused by the unique way that data is presented in tables. Tables are organized in a way that allows users to quickly look up and compare specific data points \citep{few2004show}. Furthermore, tables are effective at handling large data densities at a single time, allowing users to navigate large data sets \cite{tufte1983visual}. 

    \item \textbf{Charts } were the second preferred data output, with 36.32\% of responses preferring charts. Charts are especially good at visualizing data in a way that makes it more accessible to more people. This is illustrated in \cite{cleveland1984graphical} where the article explains how different charts are especially effective at showing trends and change over time. Furthermore, \cite{heer2010tour} explains how users can use charts and graphs to get enhance comprehension, while also simplifying the data.

    \item \textbf{Text } was the data output method that users preferred the least with 21.97\% of responses indicating that they preferred text outputs. This suggests that text was the least desirable output, as the preference was not as strong as either charts or tables. In terms of data, text limits the amount of data that can be shown at a single time \cite{card1999readings} which can make reading and understanding large data sets harder to digest. This could potentially be why users selected text as their least preferred output method.

\end{enumerate}

While applicable on their own, these results also serve as a baseline for RQ2 and RQ3. These results will act as a control when we compare and introduce user characteristics like a user's age, experience with data analysis, etc. Doing so will help us compare these results with no added demographic variables to results with demographic variables.

% Figure 5: Answer by Data Visualization Experience
\begin{figure}[t] 
\centering
\vspace{-2mm}
\includegraphics[width=.83\linewidth]{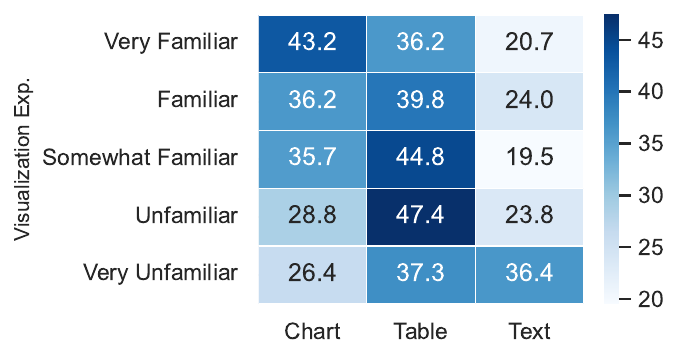}
\vspace{-3mm}
\caption{%
\textbf{User preference by Data Visualization Experience}: This shows the data preferences of respondents based on their data vis experience, specifically comparing charts, tables, and text outputs.}
\label{fig:DataViz}
\vspace{-3mm}
\end{figure}

\subsection{RQ2: User Characteristics}
RQ2 asks: \textit{Are there certain personal user characteristics that correlate with the users' preference to see a chart, table, or text?}

\subsubsection{Findings Summary}
After conducting the aforementioned study and analyzing the results, we were able to make significant findings pertaining to the relationship between personal user characteristics and their preferred data output. In short, we found that a user's familiarity with data visualization, familiarity with data analysis, and their age all influenced their data output preferences to some degree. For one, in both tests run for data analysis and visualization, we found that those with more experience in either preferred charts. Meanwhile, those with less experience were more drawn to tables. Similarly, in terms of age, younger respondents were more inclined to favor charts, while older respondents had a bias for tables. 

\medskip\noindent\textbf{Analysis Details:}
 After we got results from the respondents, we organized the respondents based on their personal characteristics; specifically looking at their familiarity with data analysis and visualizations and their age. We organized the findings into heatmaps that compared the user characteristic (y-axis) with the different data outputs. Unless otherwise mentioned, the heatmaps are normalized row-wise. 
 A p-value that is less than 0.05 shows that a statistically significant association exists and that the null hypothesis can be rejected.  The p-values revealed that there were highly significant associations between user characteristics and user preferences.

\subsubsection{H2a: Influence of Data Visualization Experience on Preference}

\textit{\textbf{H2a:} Respondents with more data visualization experience will prefer charts, while users with less experience will show a stronger preference towards tables and text.}

\textbf{Figure \ref{fig:DataViz}} shows the relationship between a user's familiarity with data visualizations and their data output preferences. As seen in the figure, participants who referred to themselves as "very familiar" with data visualizations had the strongest preference for charts (43.2\%). As experience with data visualization decreased, so did the preference for charts, with only 26.4\% of very unfamiliar respondents preferring charts. Similarly, preference for text also grew stronger as familiarity with data visualization waned, with respondents who were "very unfamiliar" preferring text at 36.4\% and tables at 37.3\%.

\textbf{Analysis:} The hypothesis that respondents with more data visualization experience will prefer charts over tables and text \textbf{\textit{was supported}} ( p < 0.01).
% , Cramer’s V = 0.0871). 
The values outputted by the study and the statistical tests both support H2a, signifying that users who are more familiar with data visualizations have a stronger preference towards charts. Charts are often only useful to those with a certain level of data visualization literacy, which could explain this trend. Similarly, the converse can be said for the stronger preference for tables and charts for those with less familiarity. Those with less familiarity might prefer these outputs as they are easier to understand with less data visualization experience. Looking at Figure \ref{fig:DataViz} from a high level, there is almost a diagonal that forms from top left to bottom right. This diagonal illustrates that as familiarity wanes, so does the preference for chart. Conversely, the preference for table gets stronger as familiarity wanes, with the "very unfamiliar" row being the main outlier.

% Figure 1: Answer by Data Analysis Experience
\begin{figure}[t]
\vspace{-2mm}
\centering
\includegraphics[width=.86\linewidth]
{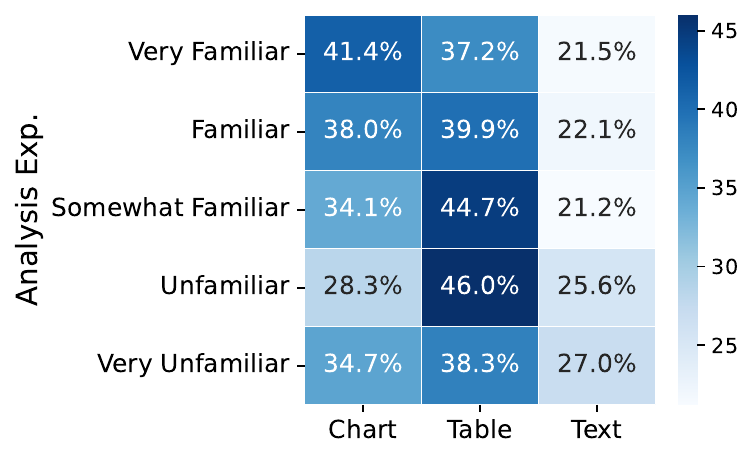}
\vspace{-2mm}
\caption{% 
\textbf{User preference by Data Analysis Experience:}
This shows the data preferences of respondents based on their data analysis experience, specifically comparing charts, tables, and text outputs. 
}
\vspace{-4mm}
\label{fig:DataAn}
\end{figure}

\subsubsection{H2b: Influence of Data Analysis Experience on Preference}

\textit{\textbf{H2b:} Respondents with more data analysis experience will also prefer charts over text and table outputs.}

\textbf{Figure \ref{fig:DataAn}} shows the relationship between a contingency table that compares a user's familiarity with data analysis to whichever data output they prefer. In creating this table, we found that data analysis experience is strongly associated with data output preference. More specifically, users who identified themselves as being very familiar with data analysis preferred charts at 41.4\%. Meanwhile, those with lower familiarity with data analysis had a stronger bias towards tables with users who identified themselves as unfamiliar and very unfamiliar preferring tables at a rate of 46.0\% and 38.3\% respectively. Finally, text output was always the least preferred but it grew marginally as familiarity waned.

\textbf{Analysis:} 
The hypothesis that respondents with more data analysis experience will prefer charts over tables and text \textbf{\textit{was supported}} ( p < 0.01). The results from our comparison of a user's data preference output with their data analysis experience are not too dissimilar to the results from the data visualization experience part of the study. For example, users with the most experience with data Analysis showed a strong preference for charts, with the preference dropping with a user's familiarity. However, the difference is that users with the least amount of familiarity with data analysis put an end to the trends in both the chart and table rows. Once again, the preference for charts among the most familiar could be because of the extra level of data literacy that is required to understand charts. Overall, this table shows that there is an association between the two variables. When designing LLMs, designers and developers can use this information to create a better user experience for their users.

% Normalized Rows
\begin{figure}[ht]
    \centering
    \includegraphics[width=1.0\linewidth]{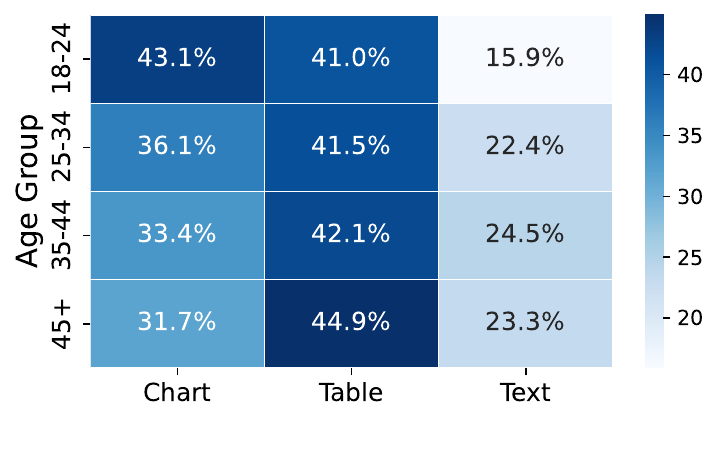}
    \caption{Comparison of the users' age to their preference in terms of whether they prefer the answer to be shown to them as a chart, table, or text.
    }
    \label{fig:age}
\end{figure}

\subsubsection{H2c: Influence of Age on Preference}

\textit{\textbf{H2c:} In terms of age, younger respondents will show a stronger preference towards charts, while older respondents will prefer tables and text.}

\textbf{Figure \ref{fig:age}} shows the relationship between a user's age range to what data output they prefer. In doing so, we found that different age groups prefer different data output methods. For one, younger users aged 18-24 showed the strongest bias toward charts at 43.1\%. On the other hand, the preference for tables increased with age with users 45 and older showing a preference for tables. Finally, text was the least preferred data output across all ages with 18-24 year olds preferring it the least at 15.9\%.

\textbf{Analysis:} The hypothesis that younger respondents will prefer charts over tables and text while older respondents would prefer the opposite \textbf{\textit{was supported}} (p < 0.01). The data outputted in the contingency table in Figure \ref{fig:age} and the p-value show that age is strongly associated with a user's data output method of choice. As mentioned, younger user's seem to prefer charts the most, but the interesting part is that this preference for charts seems to steadily drop with age, with the biggest drop coming between the 18-24 group and the 25-34 group. According to \cite{mladkova2017learning, yaru2024investigation, mellman2020getting}, younger generations prefer receiving information in easier-to-understand snippets, than larger sets of texts. Given this, it makes sense that this age group has a stronger preference for charts, with there also being a steady decline in chart preferences with each age group. Conversely, table preferences typically increase as participants get older. All in all, this data can be used when designing user experiences as it shows that a user's primary data output preference may change with age.

\subsection{RQ3: Work Experience}
RQ3 asks: \textit{Does a respondent's role at work or industry they work in correlate with their preference to see a visualization, table, or text?}

\subsubsection{Findings Summary}

\begin{figure}[t]
\centering
\includegraphics[width=1\linewidth]{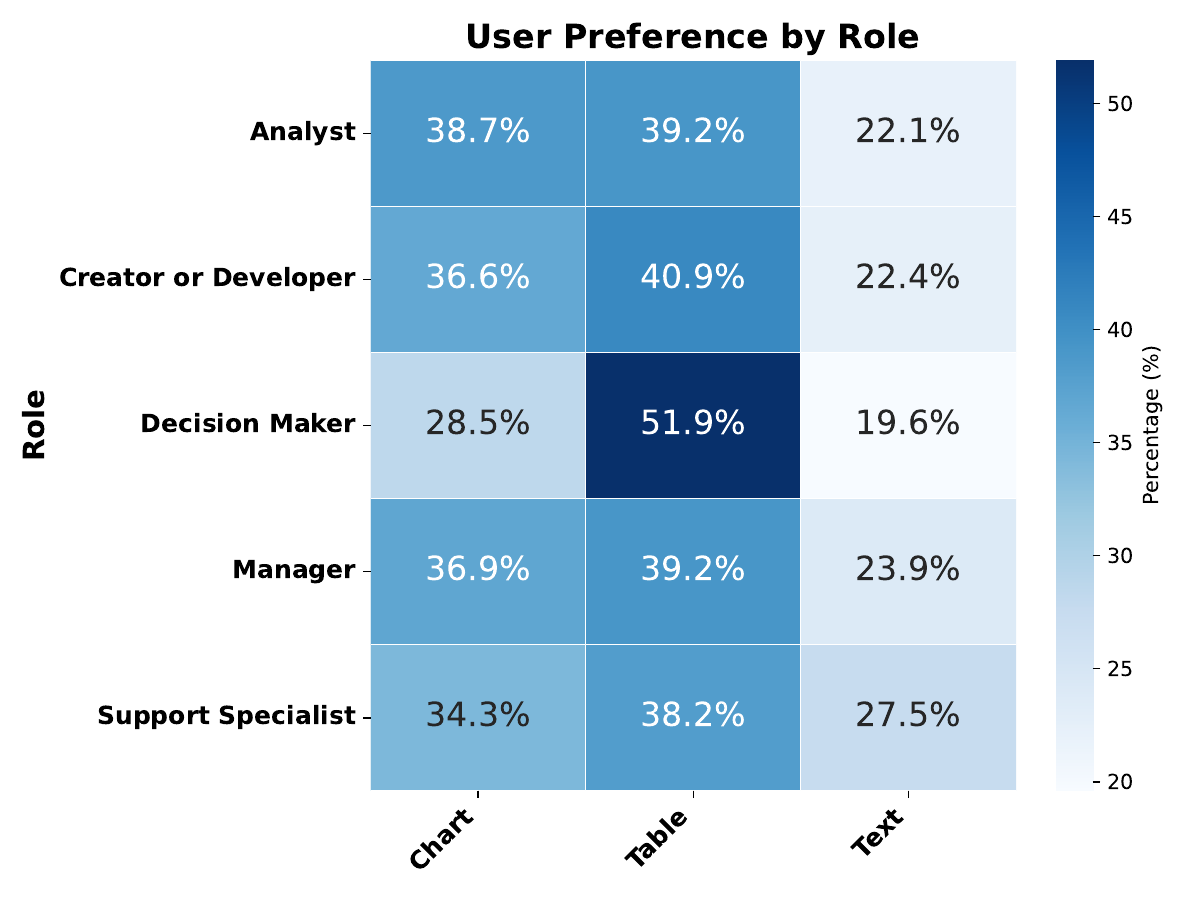}
\caption{%
\textbf{User preference by Role}: This shows the data preferences of respondents based on their work role, specifically comparing charts, tables, and text outputs. 
}\label{fig:role}
\vspace{-5mm}
\end{figure}

\begin{figure}[t]
\centering
\includegraphics[width=1\linewidth]{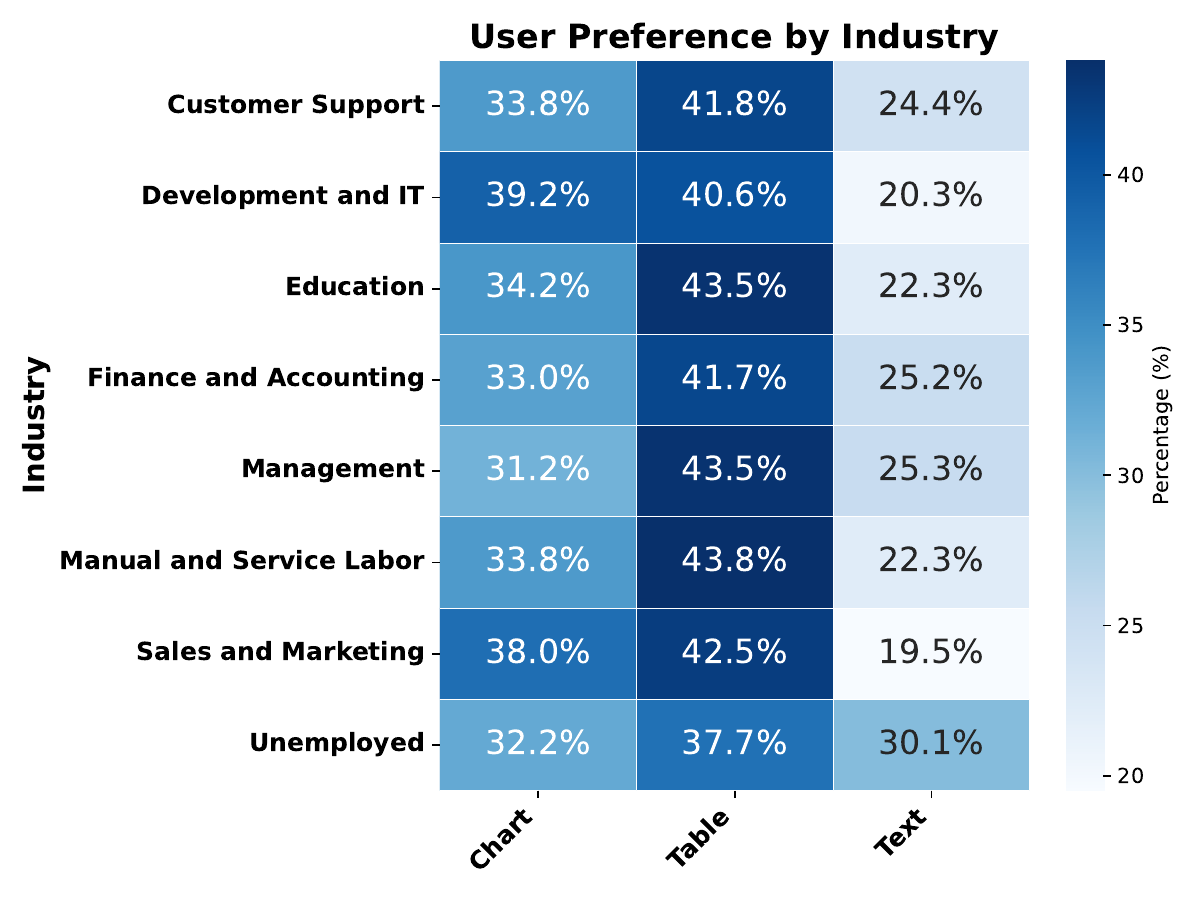}
\caption{%
\textbf{User preference by Industry}: This shows the data preferences of respondents based on their work industry, specifically comparing charts, tables, and text outputs
} \label{fig:industry}
\end{figure}

After conducting a study comparing the influence of a user's work experience on their data output preferences, we concluded that there are highly significant associations between the two. Specifically in terms of roles, we found that those in decision-maker roles strongly preferred tables, while analysts had the strongest preference for charts among the group (Fig. \ref{fig:role}). Meanwhile, in terms of a user's industry, there was a lot of variation but industries like development and IT, and Sales and Marketing preferred charts more than other industries (Fig. \ref{fig:industry}).

\textbf{Analysis Details:}
 After we got results from the respondents, we organized the respondents based on their work experiences; specifically looking at the role they played at work and the industry they worked in. Given these findings, we used p-values as forms of statistical analysis. The p-values revealed highly significant associations between the users' work experiences and the user data output preferences. 
 % Still, the Cramer's V and Chi-Square Tests revealed these associations to be weak.

% Figure 7: Answer by Role
\vspace{-3mm}
\subsubsection{H3a: Influence of Role on Preference}
\textit{\textbf{H3a:} Respondents will prefer different data outputs based on the role they play at work, with more presentation-oriented roles preferring charts.}

\textbf{Figure \ref{fig:role}} shows the relationship between a user's role at work with their data output preferences. In general, the results show that the data preferences are significantly different depending on a user's role. For example, for those who identified as analysts, charts were the most preferred data output method (38.7\%). On the other hand, decision makers were the group that preferred charts the least at 28.5\%, but preferred tables 51.9\% of the time. Finally, support specialists had the highest bias for text at 27.5\% of responses.

\textbf{Analysis:} The hypothesis that respondents with more presentation oriented workers will prefer charts over tables and text \textbf{\textit{was not supported}} ( p < 0.01). For the most part, each role had a stronger preference for tables than they did for charts, even if for some it was not by much. Considering that each role has a varied preference percentage breakdown and that the P-value is p < 0.01, there is a strong indication that there is a significant association between data preference and role. 
From these results, it can be concluded that LLMs could use a user's work role to influence what data output they use. For example, if a user is marked as a decision maker, it may make sense to show them a table given that respondents preferred tables 51.0\% of the time. Furthermore, an LLM might also want to give more weight to charts for analysts as they preferred charts 38.7\% of the time. Given all of this information, LLMs have the opportunity to be more personalized by incorporating data like this that presents data based on a user's persona.

% Figure 8: Answer by Industry

\subsubsection{H3b: Influence of Industry on Preference}
\textit{\textbf{H3b:} Respondents will prefer different data outputs based on the industry they work in, with more technical industries preferring tables and text.}

\textbf{Figure \ref{fig:industry}}  displays the relationship between the industry a user works in and whether they prefer data outputted as a table, chart, or text. From this chart, it is clear that respondents in the Development and IT industry had the highest preference for charts at 39.2\%. Meanwhile, industries like finance and accounting have a stronger preference for tables at 43.5\%. Finally, text was most strongly preferred by unemployed respondents at 30.1\%, suggesting that they prefer a narrative with their data.

\textbf{Analysis:} The findings \textit{\textbf{support}} (p < 0.01) our hypothesis as respondents preferred different data outputs based on the industries they worked in. Even more so, technical fields like Development and IT have a stronger preference for charts. This could potentially be because they are more efficient at conveying trends \citep{tufte1983visual}. Similarly, those in the Finance and Accounting industries preferred tables, suggesting that they may have wanted to look at many data points and potentially compare them \citep{few2004show}. Developers of LLMs can use this information to make their systems more responsive to users in a wide array of industries.

% Figure 3: Normalized Rows
\begin{figure}[ht]
    \centering
    \vspace{-2mm}
    \includegraphics[width=.85\linewidth]{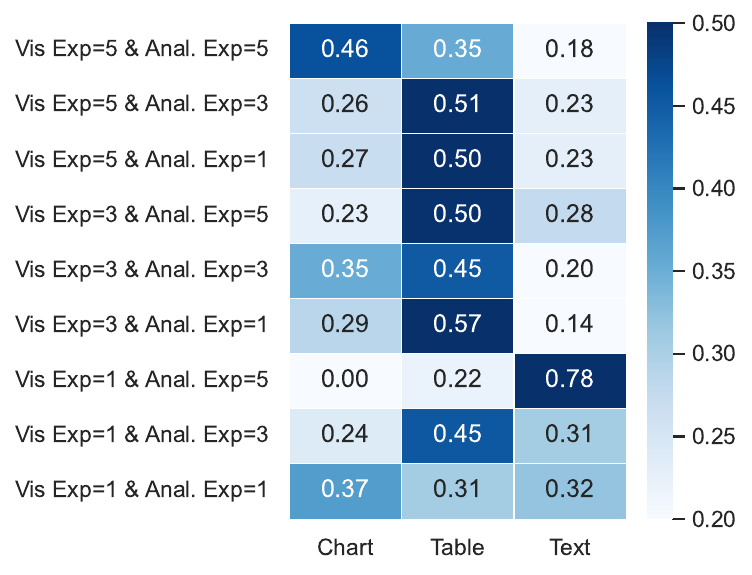}
    \vspace{-2mm}
    \caption{When users are highly experienced in both visualization and data analysis, they tend to prefer visualizations for answering questions. However, users who are unfamiliar with visualizations but experienced in data analysis lean towards text-based answers, while novices in both fields show a slight preference for visualizations over tables and text.
    }
    \label{fig:vis-and-anal-vs-preference-norm-all}
    \vspace{-2mm}
\end{figure}
    \vspace{-2mm}
\subsubsection{Preferences When Combining User Characteristics}

Figure \ref{fig:vis-and-anal-vs-preference-norm-all} highlights how the intersection of a user's combined experience with both data analysis and visualizations influences their data output preferences. This further segmentation of the data can help gain more granular insights into how different user characteristics influence a user's preferences. 

For example, if a respondent marked that they were highly familiar with both data analysis and visualizations (data visualization experience = 5; data analysis experience = 5), then they were more likely to prefer charts (46\%) over tables (35\%) or text (18\%). However, when a respondent indicated that they were experienced in only data analysis (data visualization experience = 1; data analysis experience = 5) we found that their preference shifted heavily towards text (78\%). The shift away from visualizations makes sense as these users most likely have to use text to compensate for their lack of visualization experience. Similarly, novices in both data analysis and visualizations (data visualization experience = 1; data analysis experience = 1) both show a marginal preference for charts (37\%). However, with an increase in data analysis experience, users begin to strongly prefer tables at 45\%. 

The takeaways about user preferences become more substantial when comparing two unalike user characteristics, such as a user's role and visualization experience (Figure \ref{fig:vis-and-role-vs-preference-norm-all-groupedA}). While users may have one characteristic consistent, the difference in the other characteristic often causes a sizeable swing in their data preferences. For example, an analyst with high visualization experience prefers charts (43\%), while analysts with less experience prefer tables (47\%). All in all, the fluctuations across similar roles and similar visualization experiences produce the finding that even the slightest change in user characteristics can influence their preferences.

These particular data points underscore that not all users with the same user characteristics have the same preferences. The preferences do not exist in a vacuum and often are dependent on other user characteristics. For this reason, LLMs and other data tools need to be able to dynamically adjust to a combination of user characteristics to best meet the needs of the user.

\vspace{-2mm}
\subsection{RQ4: LLM Preference Predictions}
\textit{\textbf{RQ4:} Can LLMs be used to predict whether a question should be answered with a visualization, data table, or text?}

In other words, is there alignment on this task between what humans actually prefer and the inferences generated by the LLM?
Notably, this question is of fundamental importance since if this holds, then LLMs can be used to infer how a question should be answered for a specific user.

\vspace{-2mm}

\subsubsection{H4a: LLM Alignment with Humans}
\textit{\textbf{H4a:} Using an LLM without user-specific personalization will perform poorly on predicting whether a user should be answered with a chart, table, or text.}

To answer this question, we used the approach shown in Figure~\ref{fig:basic-non-personalized-LLM-approach}.
For the LLM we used GPT4o (gpt-4o-2024-05-13b).
Using this non-personalized approach to predict the answer preferred by a user for a given question, the average accuracy is 0.367.
Notably, the average accuracy of the non-personalized LLM approach is very close to what would be expected by random selection.
This finding implies that different users often have different preferences for how they want the answer to be presented for a given data analysis question.
Hence, this result is interesting and important from a personalization perspective, and leads us to the next few research questions that seek to test whether including user-specific information including their characteristics and preferences about other questions can lead to better predictive performance.

\subsubsection{H4b: Personalized LLMs}
\textit{\textbf{H4b:} Does including user-specific examples and user characteristics in the LLM improve the accuracy of the LLM in generating the preferred answer for individual users?}

\begin{figure}[h]
\centering
\vspace{-2.8mm}
\includegraphics[width=.8\linewidth]{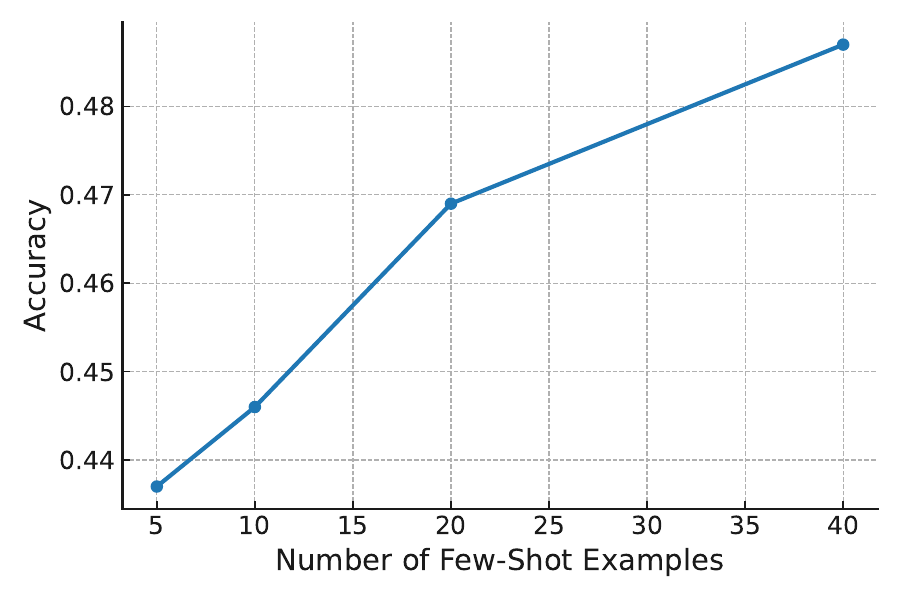}

\vspace{-2mm}
\caption{%
Our approach shows that accuracy improves as more user-specific examples are used for personalization, indicating that including additional examples of user preferences for different data analysis questions allows for more precise output. This result was based on user-specific characteristics like visualization and data analysis experience and their role.
See text for further discussion.
} \label{fig:gpt4o-few-shot-results}
\vspace{-4.1mm}
\end{figure}

To answer this question, we personalized the LLM by including the user characteristics and previous preferences that a specific user had, that is, we included questions along with how they prefer to view the answer to it (text, data table, visualization).
We provide an overview of the approach in Figure~\ref{fig:personalized-LLM-approach}.
In Figure~\ref{fig:gpt4o-few-shot-results}, we observe that the accuracy in terms of how well we personalize the responses for the specific user increases as a function of the number of user-specific examples we use for inference.
To understand the effectiveness of our approach, we also investigated the accuracy when we removed different components of our approach.
Notably, when we remove the few-shot examples, our approach achieves only 0.377 accuracy whereas when both the few-shot examples and user characteristics are removed, the performance decreases further to 0.367. We note that this last ablation is the case where no personalization is used since we do not include any user-specific examples (few-shot) and we do not provide any user characteristics to the model.
In comparison, we achieve an accuracy of 0.469 and 0.487 when 20 and 40-shots are used by the model, respectively.

As an aside, we also investigated GPT3.5 using 40-shot user-specific examples with visualization experience, data analysis experience and the users' role.
For this model, we achieved an accuracy of 0.441 compared to 0.487 using the GPT4o model.

\begin{figure}[!htbp]
\begin{formal}
\small
\textit{\tt Given the data analytics question below, along with the list of user characteristics (preferences indicated by the user), and list of questions and responses for the user, please select how the answer to the question should be presented (e.g., Table, Text, Chart) for the specific user with the user characteristics and the user's preferences for other questions.}\par
\medskip
\textit{\tt The possible options are:}\par
\textit{\tt * Table}\par
\textit{\tt * Text}\par
\textit{\tt * Chart}\par
\medskip
\textit{\tt Here are the user characteristics:}\par
\centerline{\textit{\tt [User Characteristics]}}\par
\medskip
\textit{\tt Here is a list of questions and preferences for how the user wanted the answer to be presented:}\par
\centerline{\textit{\tt [User-specific Few-shot Examples]}}\par
\medskip
\centerline{\textit{\tt [Question]}}
\end{formal}
\end{figure}
\vspace{-2mm}

\subsubsection{H4c: User-specific Accuracy}
\textit{\textbf{H4c:} Can the personalized LLM approach perform well for some users and worse for others? In other words, are some users easier to personalize and others more difficult.}

We also investigated the accuracy of individual users in Table~\ref{table:user-specific-performances}.
For brevity, we provided the accuracy of ten users across the different models.
We also selected a small subset of users and provided their user-specific accuracies from the various models investigated.

We also now show accuracy for a small subset of users, and this just shows that for some users, predicting how they want the answer to be is easy, but for others it is more difficult.

\section{Discussion} \label{sec:discussion}

This study focuses on how user characteristics influence a user's data output preferences, specifically conducting a user study measuring the preferences between charts, tables, and text outputs. We then synthesized and presented the results, and in this section, we discuss the results, examine common themes and highlight practical takeaways that can be applied to existing data tools.

\subsection{General Preferences}
Research question 1 (RQ1) looked at the issue from a bird's eye view and established the framework for later RQs, but also was significant in its own right in that it provided insight into general preferences. We hypothesized that charts would be the most preferred, but users actually generally preferred tables 41\% of the time. Charts were preferred at a somewhat similar rate of 36.2\% and text was the least preferred at 21.9\%. 

From this data, we can gather that tables are still preferred because of their ability to quickly display large data sets and allow the user to find and compare specific pieces of data \citep{few2004show}. Considering that charts are not far behind, the utility of charts should not be underestimated as they are useful in identifying trends \citep{tufte1983visual}. Finally, text still had a substantial amount of users saying they preferred it, which can have something to do with its straightforward nature or even its ability to tell a narrative.
\vspace{-1.5mm}

\subsection{Influence of User Characteristics and Work Experience}

The results show that user characteristics such as data visualization experience, data analysis experience, and age significantly influence data output preferences. Users with more experience in data visualization and analysis demonstrated a strong preference for charts, likely because they are more accustomed to interpreting trends and complex data \citep{tufte1983visual}. Meanwhile, those with less experience tended to prefer tables, possibly due to the need for quickly comparing and contrasting large data sets. Age also plays a role, as younger users favored charts, likely due to their familiarity with visually rich platforms, while older users were more inclined to prefer tables and text. This indicates that user characteristics indeed shape the way users prefer to receive data outputs.

Work experience also greatly affects data preferences, with users’ roles and the industries they work in shaping their output preferences. Analysis-oriented roles leaned towards charts, as they offer high-level insights, while decision-makers showed a stronger preference for tables, valuing the quick and accurate information they provide. Likewise, industry differences were notable: development and IT professionals preferred both charts and tables, likely for presenting trends and precise data, while those in finance and accounting favored tables due to their need for large volumes of exact, easily comparable data. These findings emphasize the opportunity to personalize data outputs based on work roles and industries, tailoring data presentations to better suit the needs of different users based on their professional backgrounds.
\vspace{-1.5mm}
\subsection{Human Preference vs. GPT Preference}

The results from investigating RQ4 reveal that providing user-specific information helps large language models predict how users will prefer to see their data. By feeding the LLMs user data like user role, data visualization experience, age, etc., we can increase the effectiveness of the models' predictions. Specifically, the LLM performed much better as we increased the user data we provided during a few shot learning, with accuracy numbers rising from 0.367 with no user data to 0.487 with forty or so examples. Providing LLMs with few shot examples significantly increased its accuracy. Overall, from this, we can gather that feeding an LLM personal data alters an LLM's outputs.

Furthermore, comparing GPT4o with GPT3.5 revealed that using the more advanced models for predicting personalization improves accuracy. Despite the finding that feeding LLM models improves accuracy, the user-specific accuracy variance still shows that some users are harder to predict. With this in mind, LLMs still require further tuning before they can be considered fully accurate or reliable in this space. With this in mind, we summarize our findings by stating that feeding personalized information improves GPT's ability to predict user preferences, but continued work should be done to optimize models to improve accuracy.

\section{Conclusion} \label{sec:conc}

In this paper, we conducted a research survey that investigated how user characteristics and work experiences shape data output preferences. We found that each of the characteristics we studied influenced a user's data output preference in some way. For this reason, we recommend that data tools tailor their outputs to the personal characteristics of each user. Doing so will create a better user experience and is likely to increase efficiency. Additionally, we used this data to explore how effective LLMs are at predicting these user preferences. Our findings indicate that when LLMs are given no personalization information, they perform poorly. However, when the LLM is provided with user-specific information, its performance improves significantly, with accuracy increasing markedly. These findings underscore the significance of understanding a user's characteristics when creating data tools and attempting to replicate preferences when using LLMs.

\bibliographystyle{ACM-Reference-Format}
\bibliography{main}

\appendix

\section*{Appendix}
\label{sec:appendix}

\section{Related Works Cont.}
\subsection{Accessibility}
Firstly, addressing traditional disabilities like vision impairments or colorblindness is crucial for creating accessible data outputs. Those who have trouble seeing may have difficulty digesting data visualizations and prefer natural language or text \citep{lundgard2021accessible}. In a similar vein, colorblind users may also have trouble with visualizations, as they often rely heavily on discerning between colors \citep{mittelstadt2015colorcat}. Data analysis also can be difficult for neurodivergent users, as data is often overwhelming and can be cognitively complex. Given these points, there exists a need to make data accessible and output it in a way that is personalized to the needs of the individual.
\section{Study Design}

\subsection{Human Annotations}

\begin{itemize}

    \item \textbf{Title:} Output Medium Preference for Data Analytics Natural Language Questions 
    
    \item \textbf{Description}: We require output preferences for questions asked to an Analytics system. Given a question about analytics data such as “What is the total revenue last month”, please select the best output medium for the question as per the given instructions.
    
    \item \textbf{Total Questions:} We aim to have participants do 20 Questions. Justification: for this is that we may want to keep it brief for MTurkers so that they do not just click through the survey. Potential drawback: not getting a large amount of results from the same MTurker.

    \item \textbf{Fifteen Minutes:} Given that we will have 20 questions with the reader having to read the instructions, this should be an ample amount of time.
    
    \item \textbf{Should Annotators do the same questions}: No. Justification: providing different types of questions can give us a wider spread of data. However, we will use the same types of questions (Summary number, visualization, etc.) Potential Drawback: Maybe more variation.

    \item \textbf{Price:} ".40 cents" Given that we will have up to 20 questions, we may want to offer more money than usual as the survey will be longer.

    \item \textbf{Hypotheses}: 
    \begin{itemize}
        \item 1. Participants with more of a data analytics background will prefer text outputs or tables for more precise information. 
        
        \item 2. Participants with more of a managerial or decision-making role will prefer visualizations as they may be more presentation-oriented.

        \item 3. Older participants will prefer outputs with more visualizations while younger participants will prefer outputs that are primarily text.

        \item 4. There will be a strong correlation between output preference and device preference, with mobile user preferring textual outputs.

        \item 5. Participants in more technical industries (finance, IT, and Tech) will prefer charts and text outputs as they are more precise.

        \item 6. The preferences of human participants will align closely with the simulated preferences of an LLM
    \end{itemize}
\end{itemize}

\subsection{User Characteristics Questions} \label{demoQs}
\begin{itemize}
%[noitemsep]
    \item[\textbf{QA:}] How would you rate your experience with data analysis?
    \begin{itemize}
        \item Responses ranged from very unfamiliar to very familiar.
    \end{itemize}
    \item[\textbf{QB:}] How would you rate your experience level interacting with data visualizations?
    \begin{itemize}
        \item Responses ranged from not familiar or very familiar.
    \end{itemize}
    \item[\textbf{QC:}] What industry do you currently work in?
    \begin{itemize}
        \item Response examples include education, management, customer support, and other similar options.
    \end{itemize}
    \item[\textbf{QD:}] Which of the following best describes your primary role at work?
    \begin{itemize}
        \item Response examples include decision maker, analyst, manager, and other similar options.
    \end{itemize}
    \item[\textbf{QE:}] Which age range best describes you?
    \begin{itemize}
        \item Responses include age ranges from 18 to 45+.
    \end{itemize}
    \item[\textbf{QF:}] Which education level best describes you?
    \begin{itemize}
        \item Response examples include answers that range from high school to graduate school.
    \end{itemize}
    \item[\textbf{QG:}] Which gender best describes you?
    \begin{itemize}
        \item Response examples male, female, non-binary, and prefer not to answer
    \end{itemize}
    \item[\textbf{QH:}] Which device do you use most frequently for work-related tasks?
    \begin{itemize}
        \item Response examples include phones, desktop computers, and tablets.
    \end{itemize}
\end{itemize}

\subsection{Survey Questions} \label{demoQs}
Some examples of questions that were used in our survey are as follows:

\begin{itemize}%[noitemsep]
    \item Forecast the growth rate of our paid customers in the next quarter.
    \item Show me the distribution of user characteristics based on age groups.
    \item What are my top 20 attributes by highest segment count?
    \item What is the average time spent by users on the website in the last week?
    \item Display the distribution of monthly revenue from different shipping methods.
\end{itemize}

\section{Further Discussion}
\subsection{Implications for Data Tools}

Given that we have established that both user characteristics and work experiences influence data output preferences, there are significant opportunities to use these findings to design data tools and large language models (LLMs) that better meet the individual needs of users. For example, a system could identify a user as a young data analyst and could use this information to confidently display a chart. Similarly, an older decision-maker in finance or accounting could be recommended a table. Either way, there is an opportunity for LLMs and other data tools to use these insights to create a user experience, increasing satisfaction and efficiency.

We do not, however, expect this data to be the end-all-be-all for data tools. Instead, this data should be used as a foundation to be built upon. LLMs, for example, could use this foundation but continue to build upon it by gathering user preferences over time. As data and personalization continue to grow interdependent on one another, the insights from this study can continue this growth and lay the foundation for tools that better fit the needs of users.

\subsection{Influence of Work Experience}
A user's work experience, specifically the role they play at work and the industry they work in, significantly affects the data outputs they prefer. In terms of a user's role at work, we found that analysis-oriented roles have a stronger preference for charts. Meanwhile, those in decision-making roles had the strongest preference and preferred tables about 52\% of the time. From this data, we gathered that a user's role influences their data output preference. Workers who preferred charts may have liked the high-level insights provided by charts, while those who preferred tables may have appreciated the quick and accurate information they offer.

On a similar note, a user's industry often influences the data output they prefer most. Notably, those in development and IT preferred both charts and tables at a higher percentage, potentially due to the need to display and present data trends to their colleagues while also needing precise data. On the other hand, those in finance and accounting preferred tables, which aligns with their need for large amounts of precise data points that are easy to compare. Overall, we found that industries with a more visual storytelling nature tend to prefer charts, while those that require precision prefer tables.

In total, these findings support the overall thesis that there is an opportunity to display data in a way that is personalized based on a user's background. Specifically, different workers prefer to see their data displayed in various ways depending on their roles and the industry in which they work. This should come as no surprise, as different work roles have different data needs: what works well for one industry might be detrimental in another and vice versa. At a high level, these takeaways can be used to create data experiences that better cater to the individual user based on their work industry or role.

\begin{figure}[H]
    \centering
    \includegraphics[width=0.9\linewidth]{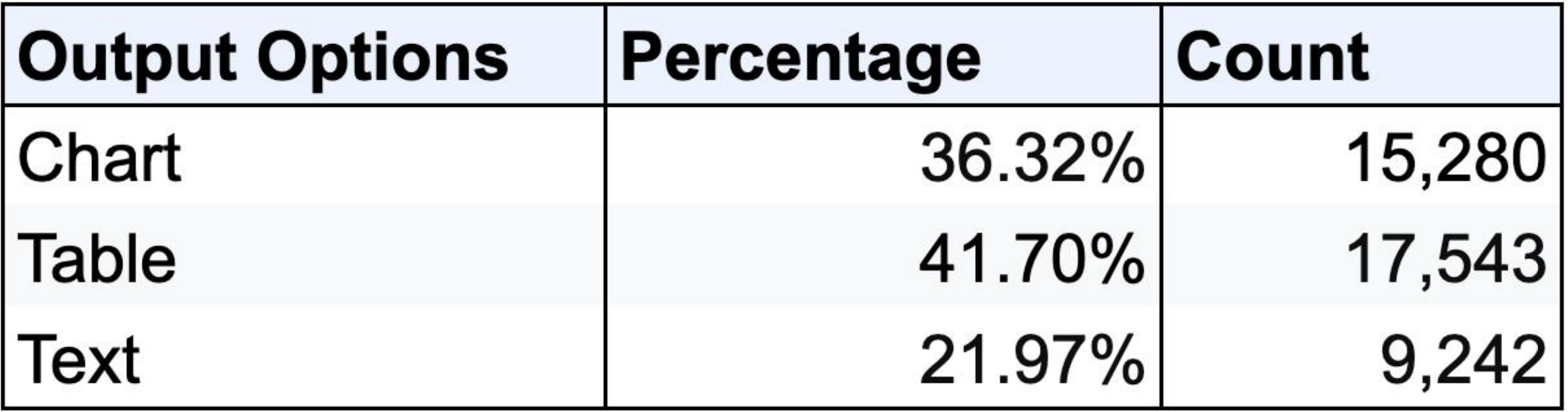}
    \caption{Ranking results for RQ1 showed that users preferred Tables at 41.7\% with Charts trailing at 36.32\%, and Text being the least preferred at 21.97\%. 
    }
    \label{fig:RQ1Chart}
\end{figure}

\begin{figure}[H]
\centering
\includegraphics[width=.99\linewidth]{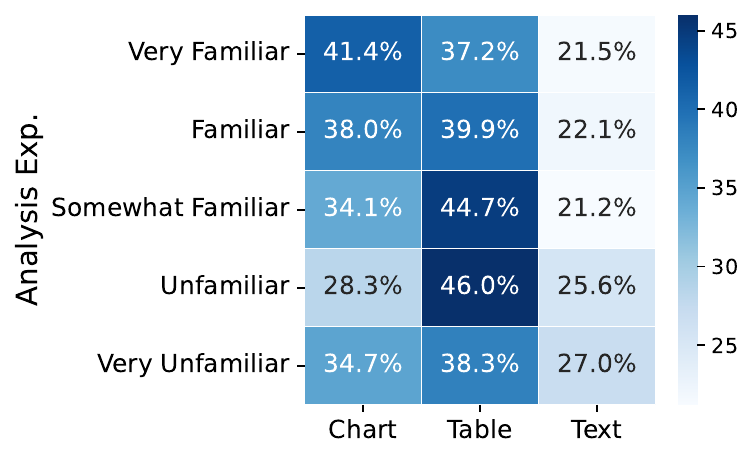}
\caption{% 
\textbf{The chart shows that if the user is more familiar with data analysis, they prefer charts more.
However, if they have less familiarity they begin to prefer tables much more.
Furthermore, users with more data analysis experience prefer text less, and the preference towards text increases as a function of their data analysis lack of experience (e.g., 21.5\% for very experienced to 27\% for very inexperienced).
}:
\textbf{User preference by Data Analysis Experience}: Answer by Data Analysis Experience:
}
\end{figure}

% Mturk Survey pics
\begin{figure}[H]
\centering
\includegraphics[width=.99\linewidth]{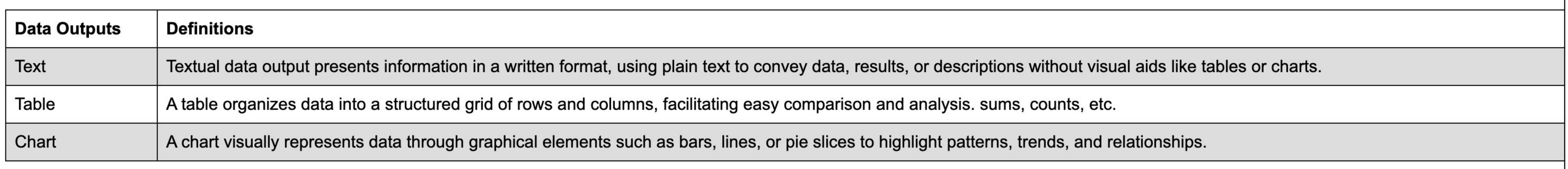}
\caption{Definitions from user survey}
\end{figure}

% Mturk Survey pics
\begin{figure}[H]
\label{ExampleQ}
\centering
\includegraphics[width=.99\linewidth]{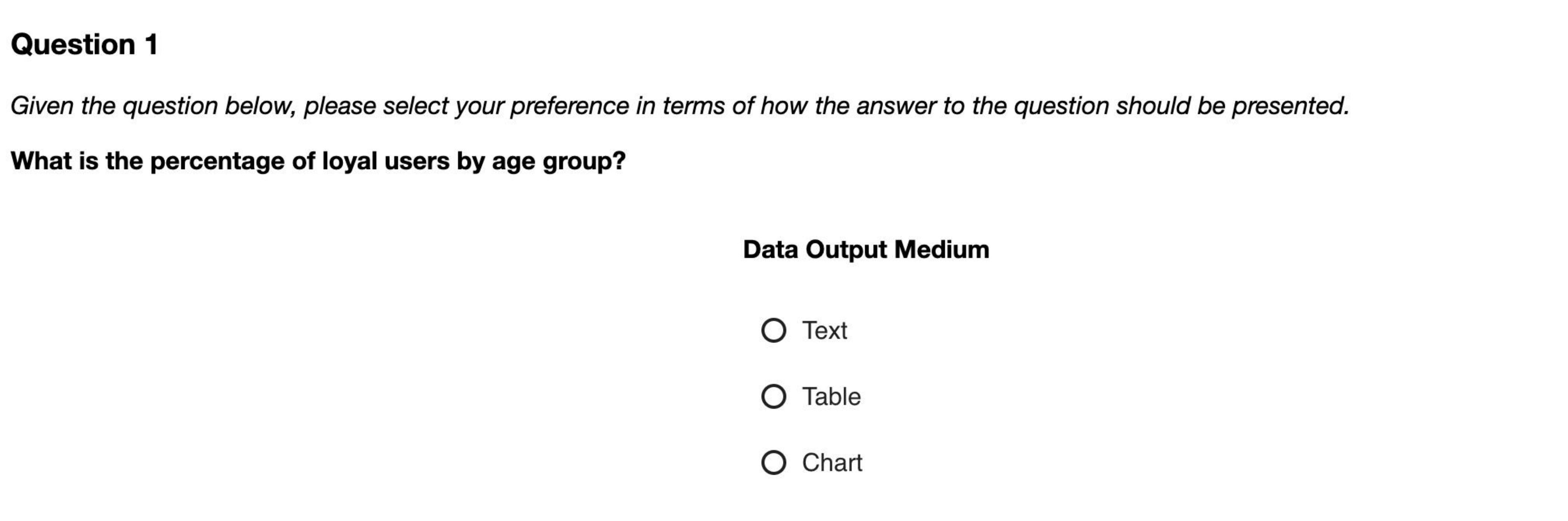}
\caption{Example question from user survey}
\end{figure}

% Mturk Survey pics
\begin{figure}[H]
\centering
\includegraphics[width=.99\linewidth]{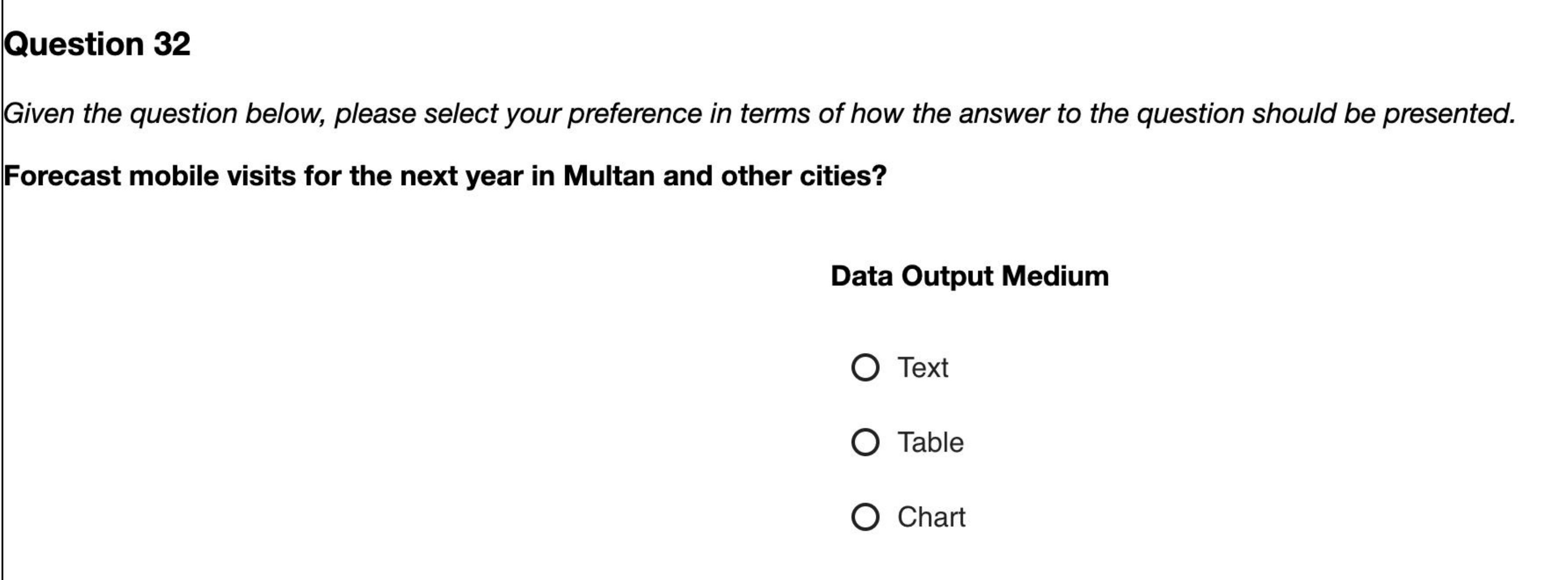}
\caption{Example question from user survey}
\end{figure}

%DemoA
\begin{figure}[H]
\centering
\includegraphics[width=.99\linewidth]{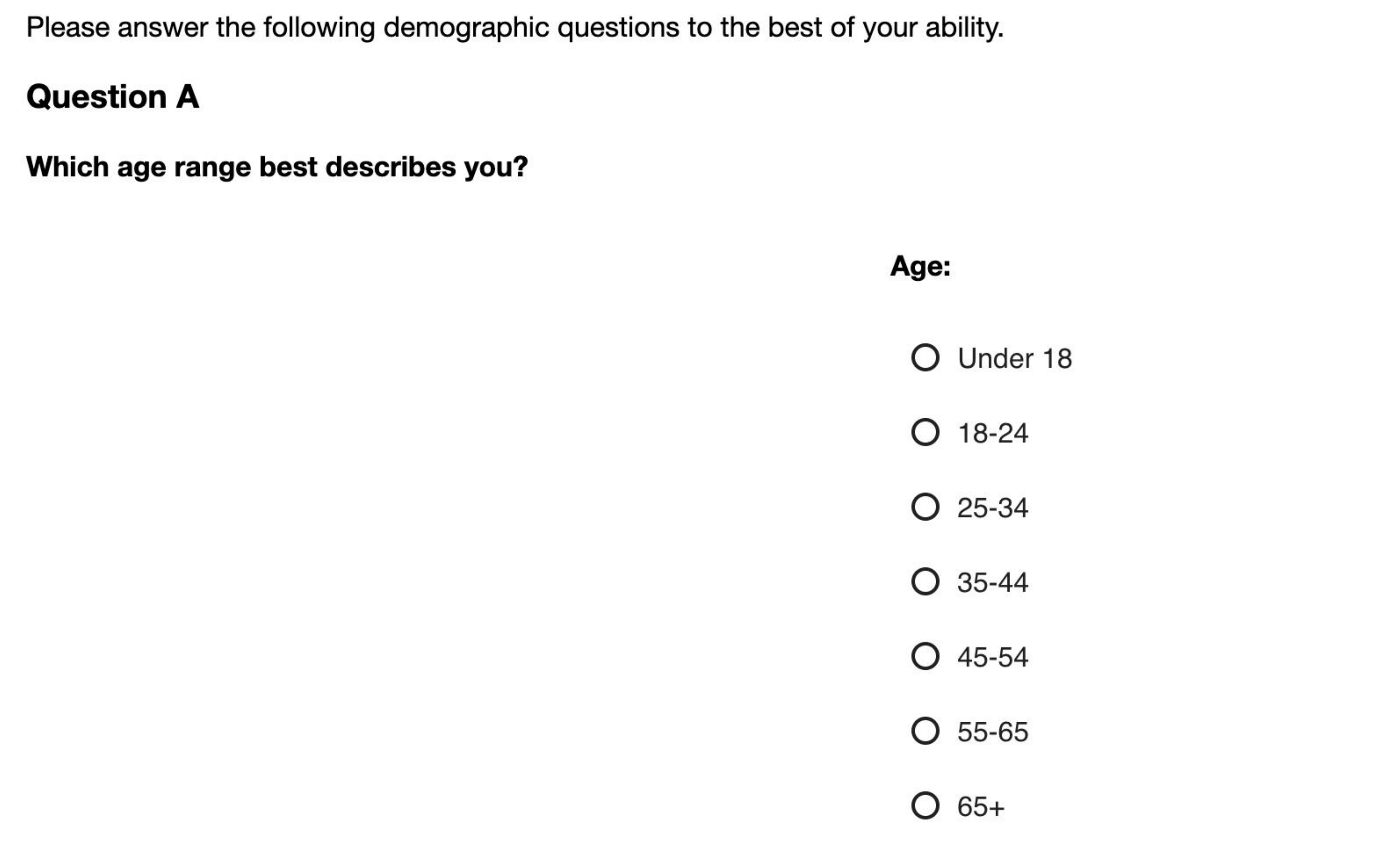}
\caption{Example question from user survey}
\end{figure}

%Democ
\begin{figure}[t]
\centering
\includegraphics[width=.99\linewidth]{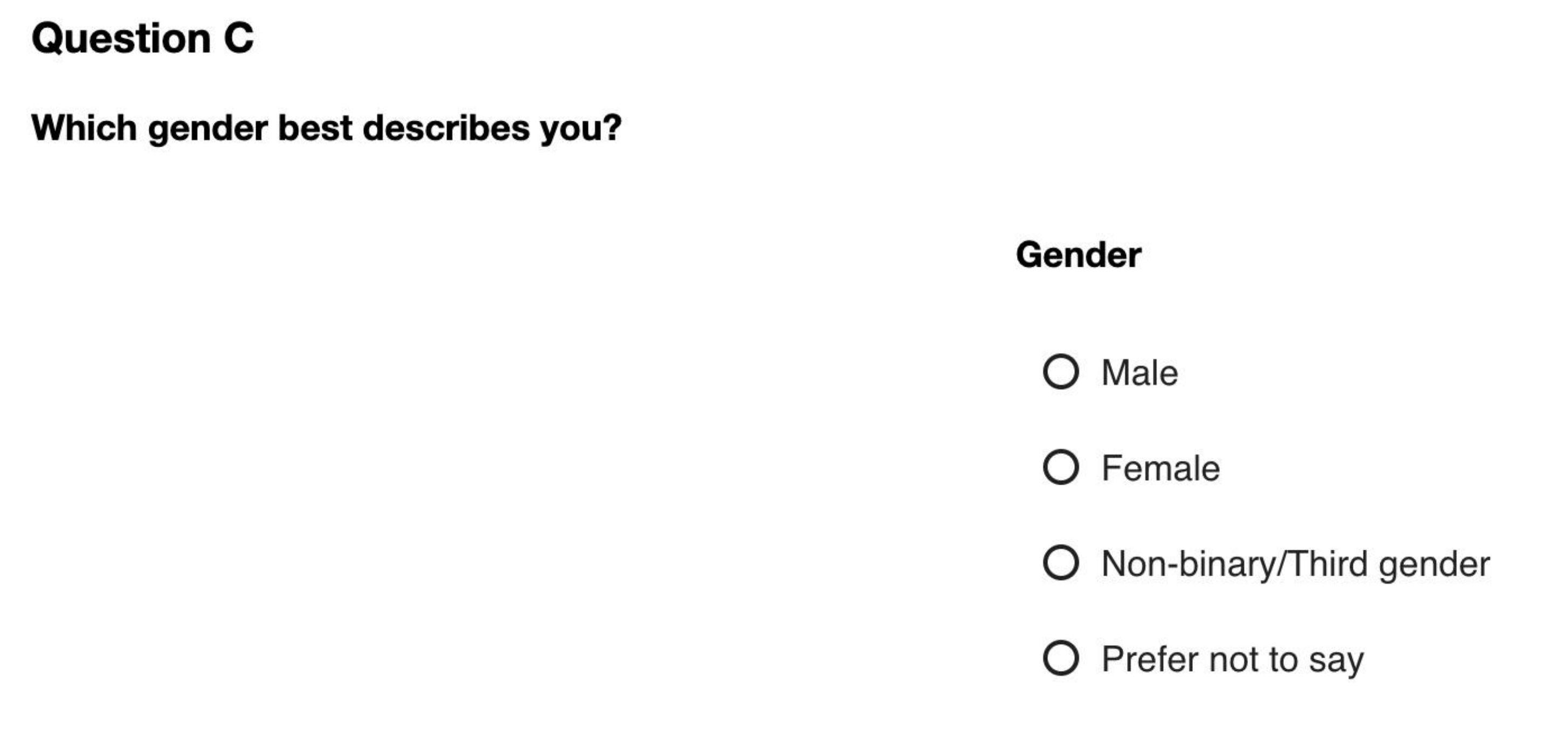}
\caption{Example question from user survey}
\end{figure}

%DemoD
\begin{figure}[t]
\centering
\includegraphics[width=.99\linewidth]{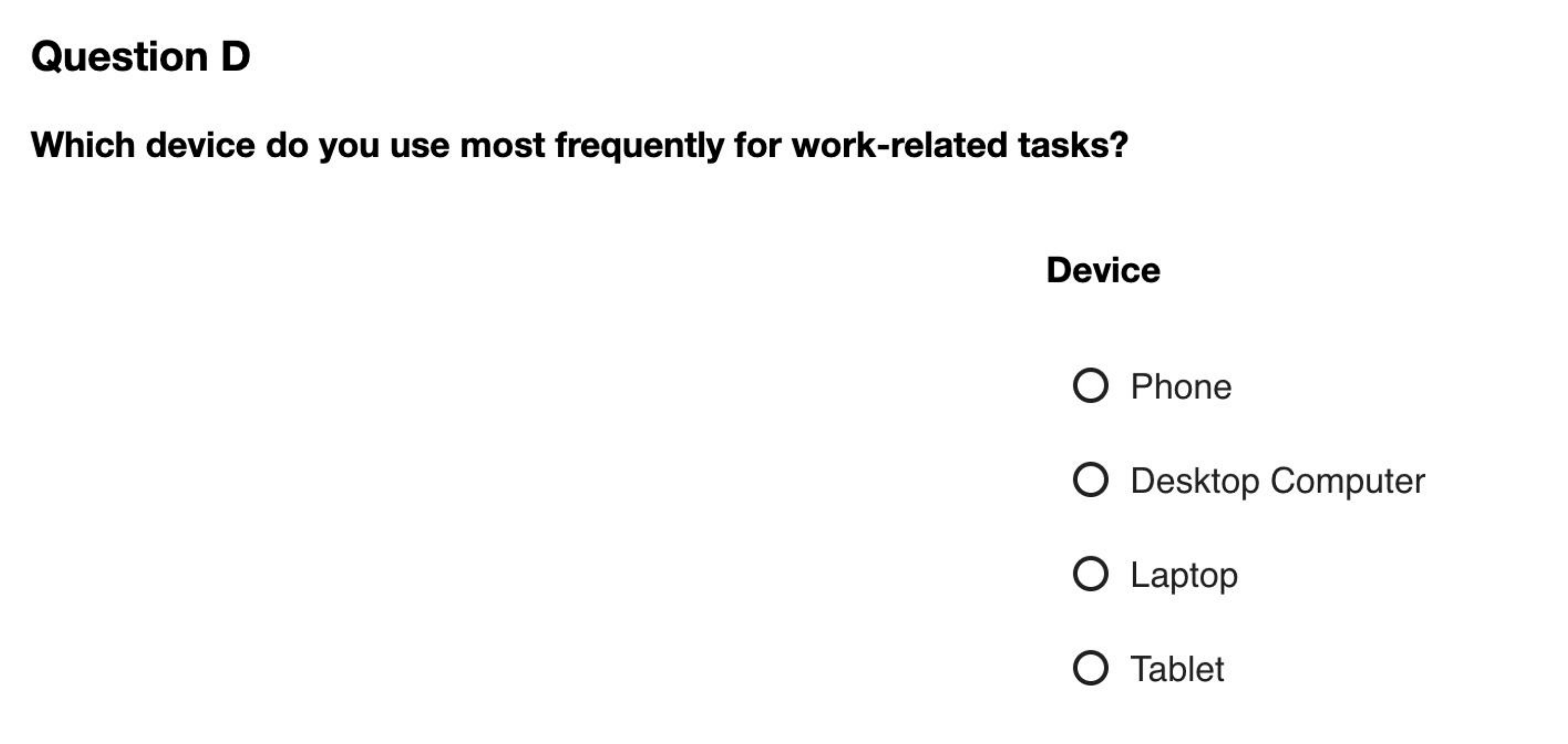}
\caption{Example question from user survey}
\end{figure}

%DemoE
\begin{figure}[t]
\centering
\includegraphics[width=.99\linewidth]{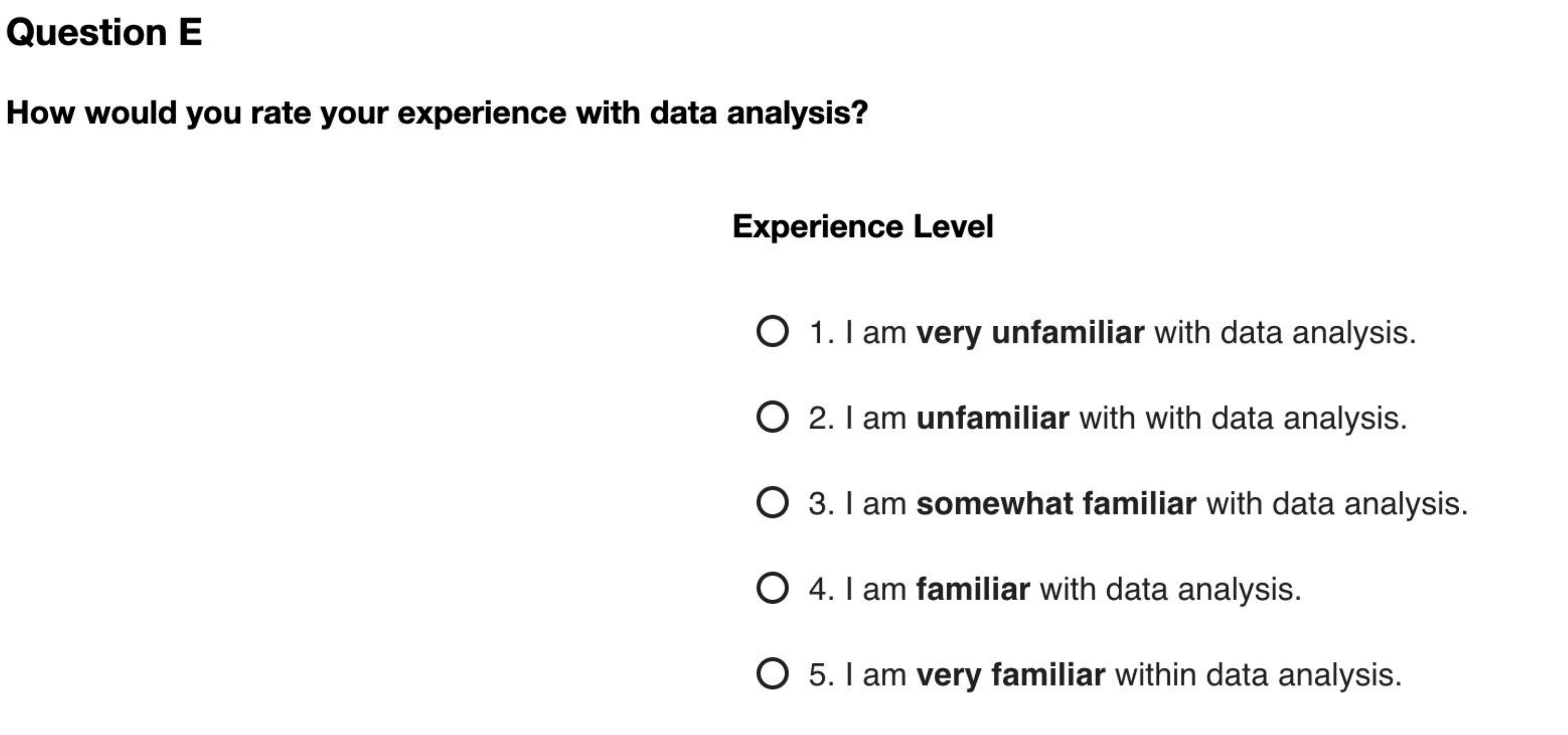}
\caption{Example question from user survey}
\end{figure}

%DemoF
\begin{figure}[t]
\centering
\includegraphics[width=.99\linewidth]{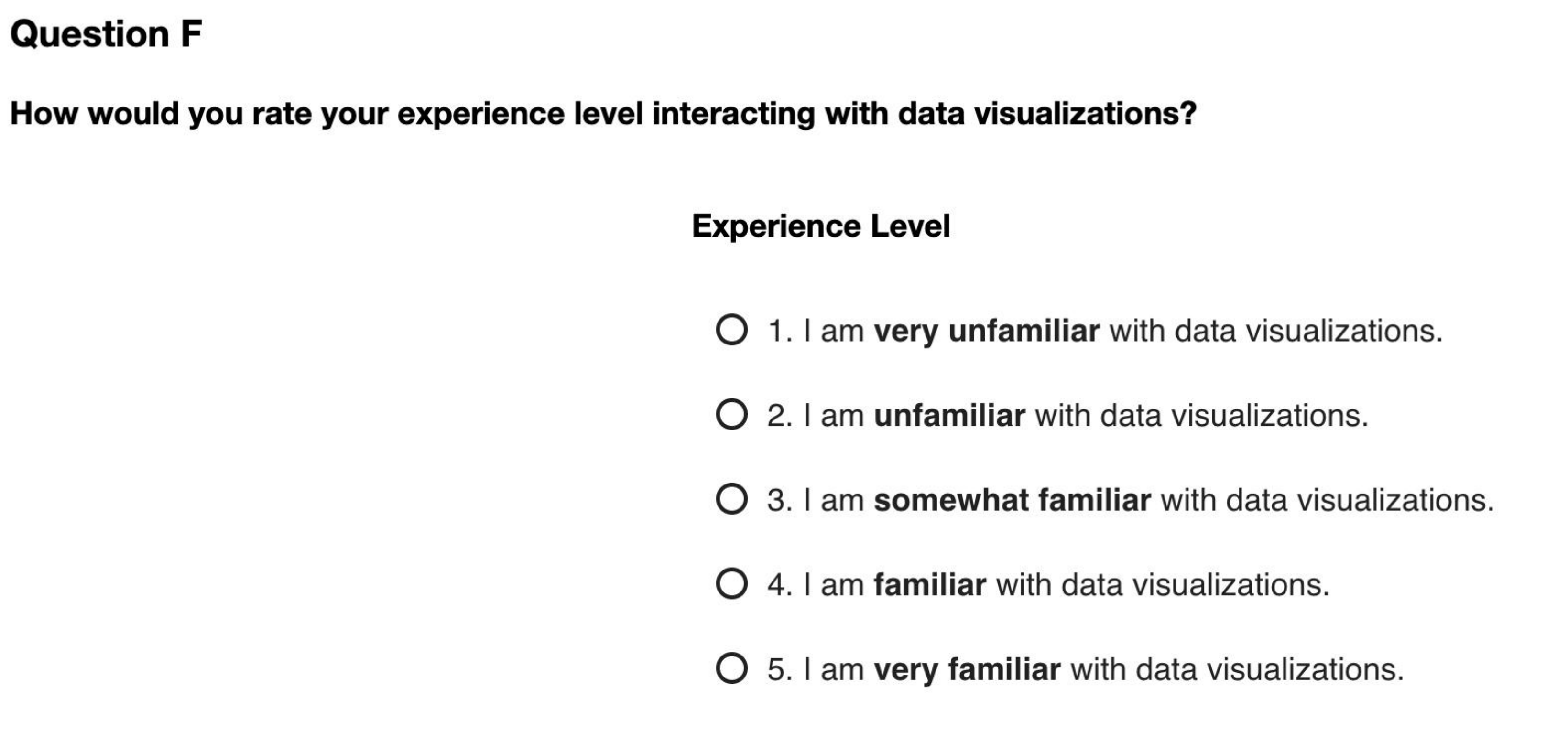}
\caption{Example question from user survey}
\end{figure}

% Figure 1: Answer by Data Analysis Experience and Data Visualization Experience
\begin{figure*}[t]
\centering
\includegraphics[width=.99\linewidth]{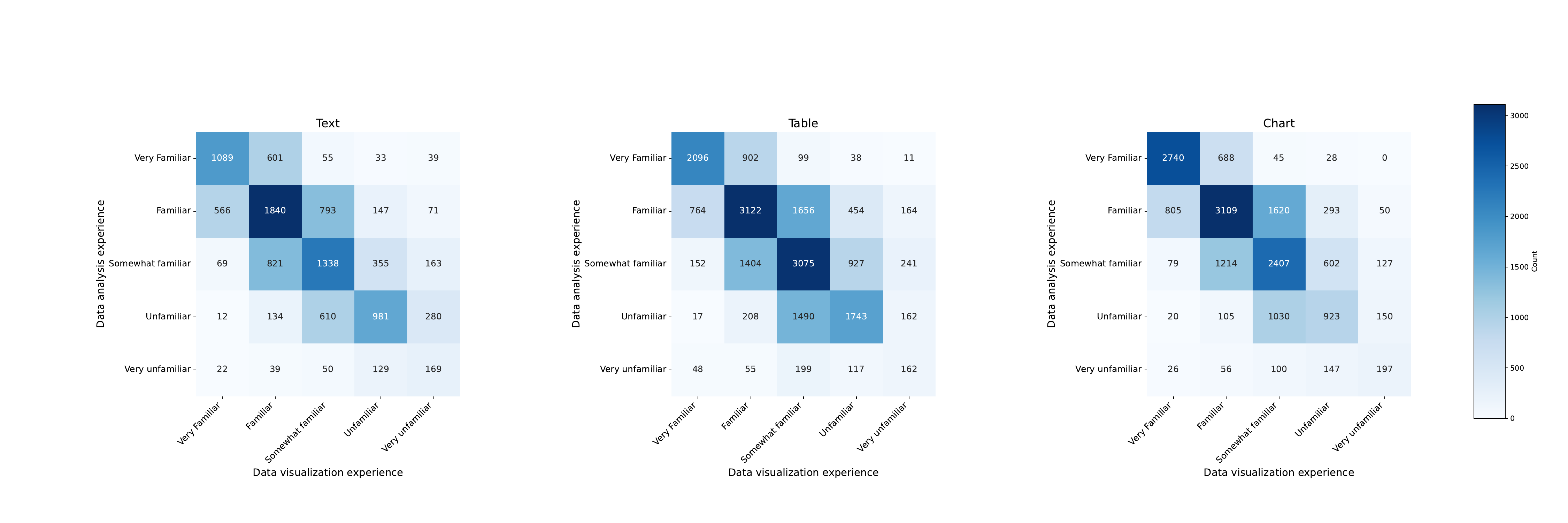}
\caption{User preferences by Data Analysis Experience and Data Visualization Experience.}
\end{figure*}

% Figure 3: Normalized Rows
\begin{figure}[ht]
    \centering
    \includegraphics[width=1.0\linewidth]{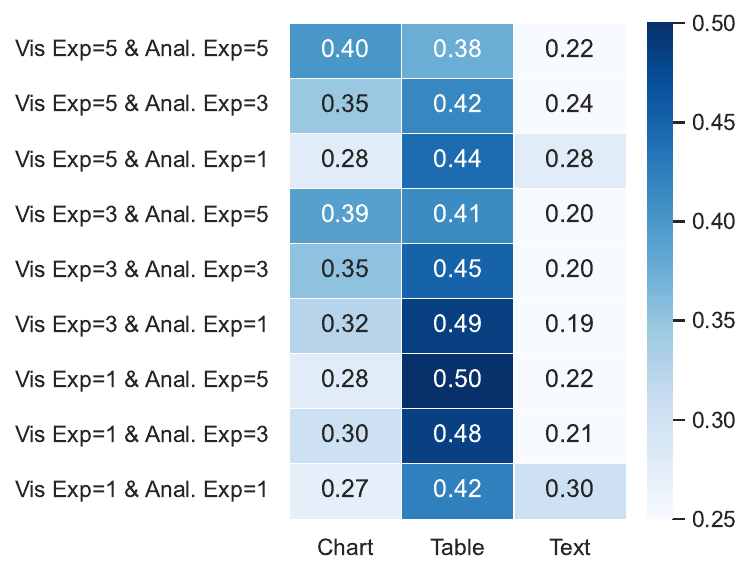}
    \caption{Comparing the users experience with visualizations and data analysis to their preference in terms of whether they prefer the answer to be shown to them as a chart, table or text.
    Note we normalize each output type (rows) and combined counts of very familiar and familiar, and very unfamiliar and unfamiliar.
    }
    \label{fig:vis-and-anal-vs-preference-norm-all-grouped}
\end{figure}

% Figure 2: Normalized Column
\begin{figure}[ht]
    \centering
    \includegraphics[width=1.0\linewidth]{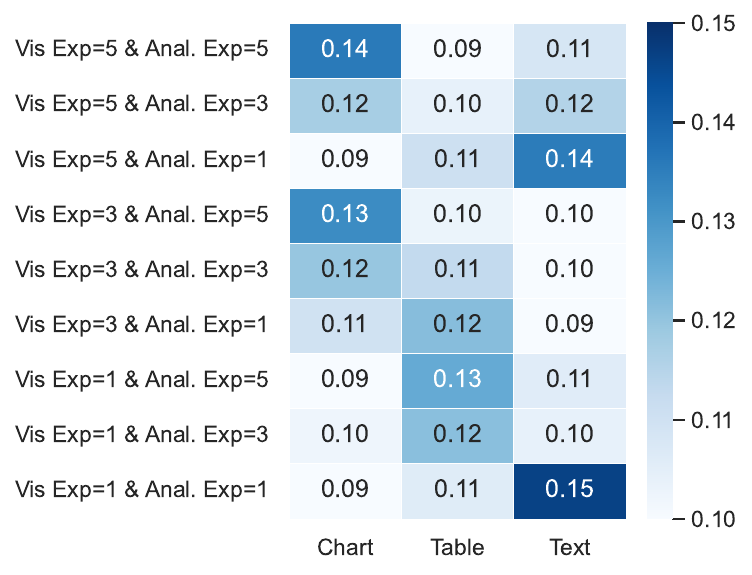}
    \caption{Comparing the users experience with visualizations and data analysis to their preference in terms of whether they prefer the answer to be shown as a chart, table or text.
    Note we normalize each output type (columns) and combined counts of very familiar and familiar, and very unfamiliar and unfamiliar.
    }
    \label{fig:vis-and-anal-vs-preference-norm-col-grouped}
\end{figure}
% }%

% Normalized Rows
\begin{figure}[ht]
    \centering
    \includegraphics[width=1.0\linewidth]{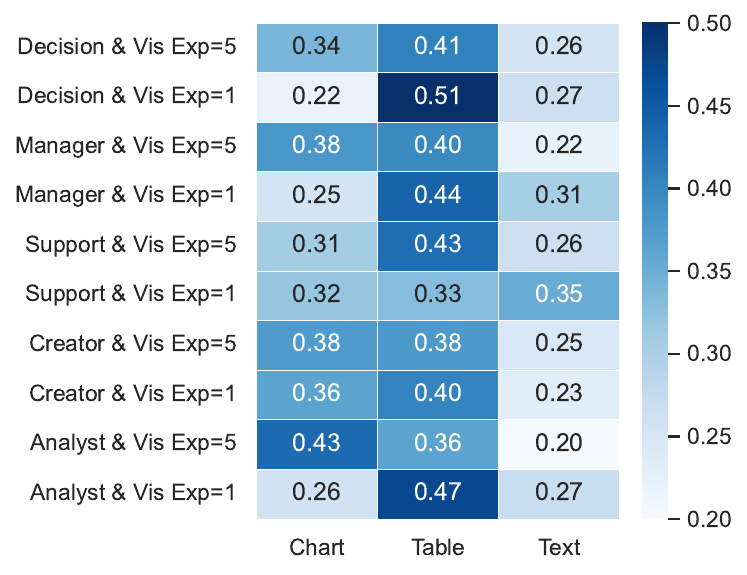}
    \caption{Comparing the users' experience with visualizations and the role of the user to their preference in terms of whether they prefer the answer to be shown to them as a chart, table, or text.
    Note we normalize each output type (rows) and combine counts of very familiar and familiar, and very unfamiliar and unfamiliar. 
    }
    \label{fig:vis-and-role-vs-preference-norm-all-groupedA}
\end{figure}

% Normalized Rows
\begin{figure}[h]
    \centering
    \includegraphics[width=1.0\linewidth]{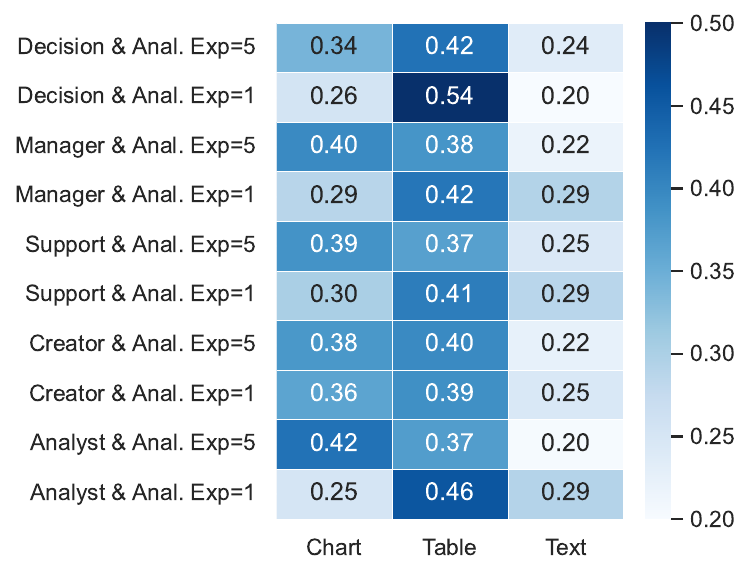}
    \caption{Comparing the users experience with data analysis and the role of the user to their preference in terms of whether they prefer the answer to be shown to them as a chart, table or text.
    Note we normalize each output type (rows) and combine counts of very familiar and familiar, and very unfamiliar and unfamiliar.
    }
    \label{fig:data-analysis-and-role-vs-preference-norm-all-grouped}
\end{figure}

% Normalized Rows
\begin{figure}[h]
    \centering
    \includegraphics[width=1.0\linewidth]{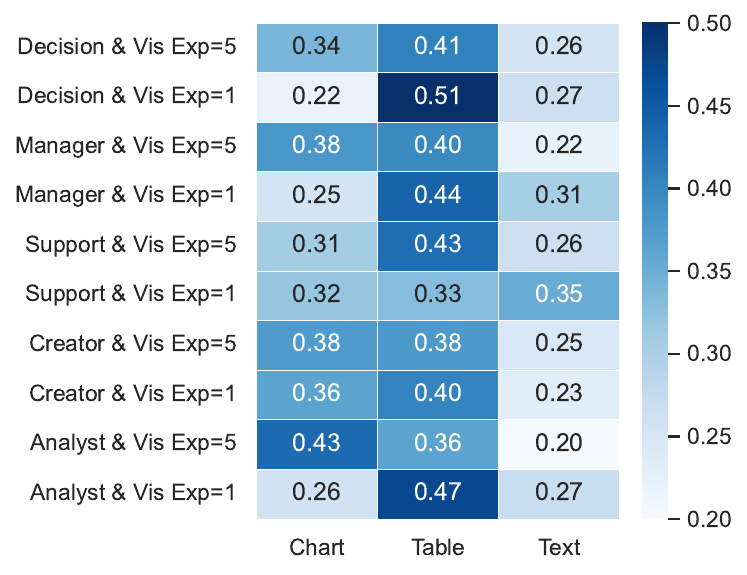}
    \caption{Comparing the users experience with visualizations and the role of the user to their preference in terms of whether they prefer the answer to be shown to them as a chart, table or text.
    Note we normalize each output type (rows) and combine counts of very familiar and familiar, and very unfamiliar and unfamiliar.
    }
    \label{fig:vis-and-role-vs-preference-norm-all-grouped}
\end{figure}

\begin{figure}[h!]
    \centering
    \begin{small}
    \begin{formal}
    \begin{quote}
    \small
    \tt
    Given the data analytics question below, along with the list of user characteristics (preferences indicated by the user), and list of questions and responses for the user, please select how the answer to the question should be presented (e.g., Table, Text, Chart) for the specific user with the user characteristics and the user’s preferences for other questions.

    \vspace{0.5em} % Adds a small vertical space instead of \\

    The possible options are:
    * Table
    * Text
    * Chart

    \vspace{0.5em} % Adds a small vertical space

    Here are the user characteristics:
    * visualization experience: somewhat familiar (3/5)
    * data analysis experience: unfamiliar (2/5)
    * role: manager

    \vspace{0.5em} % Adds a small vertical space

    Here is a list of questions and preferences for how the user wanted the answer to be presented:

    Question: What is the total count of ad clicks recorded? \\
    Answer: Text

    Question: Analyze the distribution of Click-Throughs by Age groups (20-25) in Serbia \\
    Answer: Table

    Question: Segment our user base by location (urban, suburban, rural) in the automotive industry. \\
    Answer: Text

    Question: Check in on mobile visits in July and check the overlap with loyalty level B? \\
    Answer: Table

    Question: What is the distribution of cart views across various entry pages (X, Y, Z) in Serbia? \\
    Answer: Table

    Note that:
    * Output the answer with the prefix "Answer:" on a separate line.
    * Do not include any explanations in the answers.
    * The user experience with visualization and data analysis is on a 5-point Likert scale with options from very familiar (5) to very unfamiliar (1).

    \vspace{0.5em} % Adds a small vertical space

    Question: Compare revenue for the US \\
    Answer: 
    \end{quote}
    \end{formal}
    \end{small}
    \caption{Example of how user characteristic information was fed into GPT one step at a time. This approach allowed GPT to use each user characteristic quality to add to the overall persona, resulting in answers based on the user’s characteristics.}
    \label{fig:personalized-LLM-approach}
\end{figure}

\begin{table}[h]
\centering
\small
\begin{tabular}{ccccccc}
\toprule
\textbf{Users} & \textbf{K=40} & \textbf{K=20} & \textbf{K=10} & \textbf{K=5} & \textbf{No Few-Shot} & \textbf{No Person.} \\ 
\midrule
$u_1$  & 0.80 & 0.80 & 0.75 & 0.73 & 0.54 & 0.64 \\ 
$u_2$  & 0.81 & 0.77 & 0.78 & 0.82 & 0.28 & 0.20 \\ 
$u_3$  & 0.80 & 0.53 & 0.55 & 0.42 & 0.28 & 0.18 \\ 
$u_4$  & 0.70 & 0.77 & 0.68 & 0.64 & 0.64 & 0.62 \\ 
$u_5$  & 0.70 & 0.47 & 0.38 & 0.31 & 0.30 & 0.28 \\ 
$u_6$  & 0.67 & 0.68 & 0.68 & 0.60 & 0.53 & 0.51 \\ 
$u_7$  & 0.60 & 0.40 & 0.40 & 0.36 & 0.26 & 0.26 \\ 
$u_8$  & 0.50 & 0.40 & 0.33 & 0.36 & 0.34 & 0.32 \\ 
$u_9$  & 0.40 & 0.37 & 0.33 & 0.36 & 0.34 & 0.28 \\ 
$u_{10}$ & 0.40 & 0.34 & 0.15 & 0.23 & 0.18 & 0.16 \\ 
\bottomrule
\end{tabular}
\caption{Results for a small set of users, showing the accuracy across different few-shot values and personalization approaches.}
\label{table:user-specific-performances}
\end{table}

\begin{figure}[h!]
\begin{formal}
\small
\textit{\tt Given the data analytics question below, please select how the answer to the question should be presented (e.g., Table, Text, Chart) for the specific user.}
\vspace{0.5em} % Adds a small vertical space

\textit{\tt The possible options are:}
\begin{itemize}
    \item \textit{\tt Table}
    \item \textit{\tt Text}
    \item \textit{\tt Chart}
\end{itemize}
\vspace{0.5em} % Adds a small vertical space

\centerline{\textit{\tt [Question]}}
\end{formal}
\caption{Non-personalized approach. This is the basic approach that is not personalized for a specific user.}
\label{fig:basic-non-personalized-LLM-approach}
\end{figure}

\begin{figure}[h]
\centering
\includegraphics[width=1.0\linewidth]{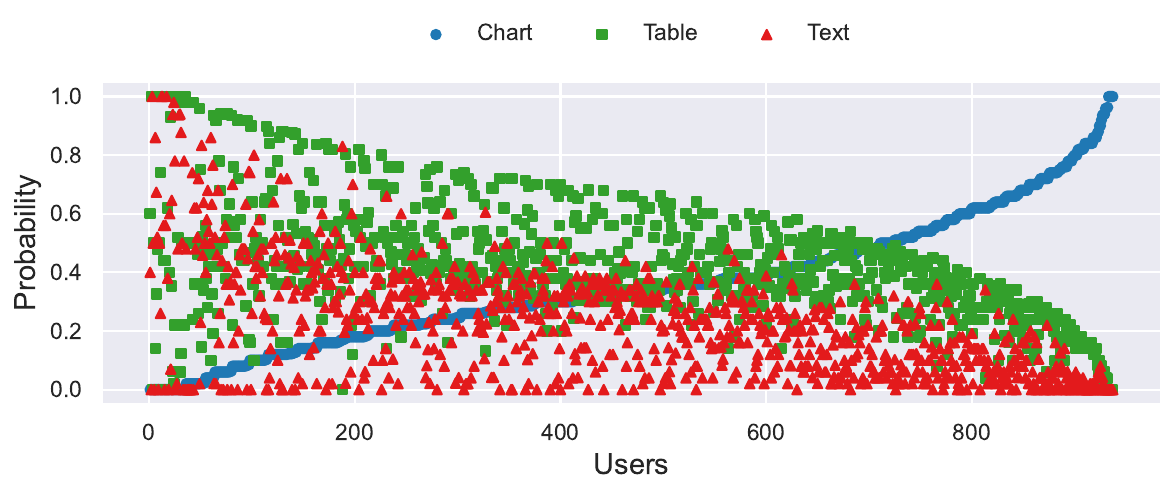}
\caption{For each user, we aggregate all their preferences and derive a distribution, shown above. The values are sorted by the probability of choosing 'chart', creating the observed curve.}
\label{fig:user-distribution}
\end{figure}

\begin{figure}[t]
\centering
\subfigure[Questions CCDF]{
\includegraphics[width=1.0\linewidth]{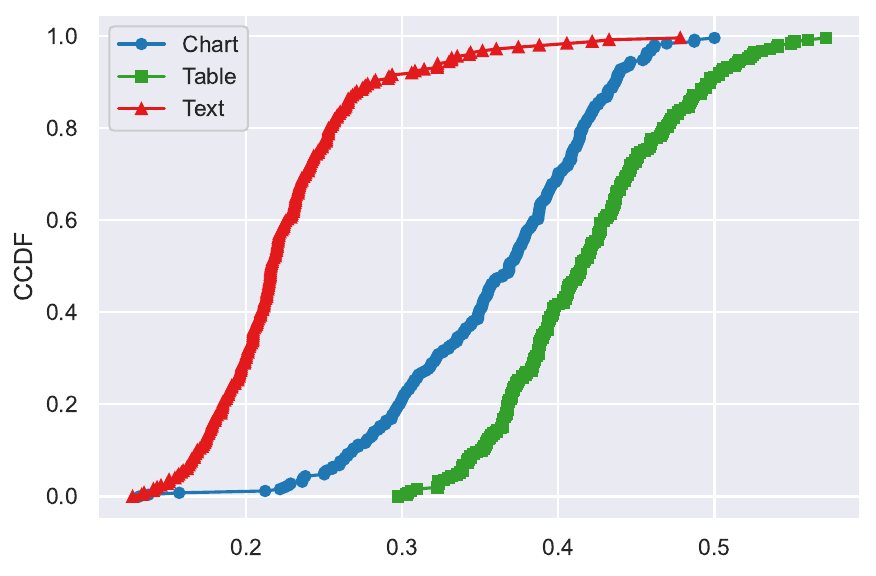}
}
\subfigure[Users CCDF]{
\includegraphics[width=1.0\linewidth]{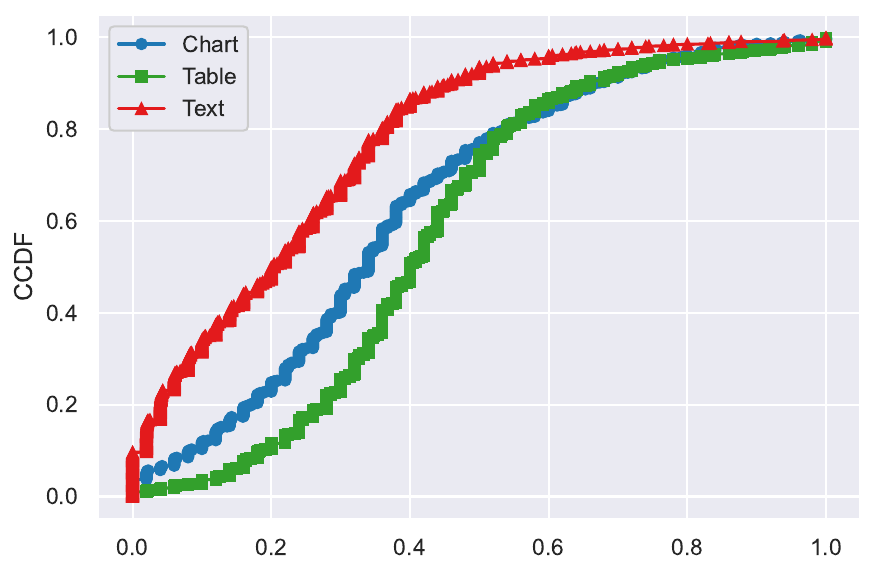}
}
\caption{%
CCDF for the questions and users.
For each question, we derive a distribution of user responses, and conversely, for each user, we derive the distribution of responses and for both compute the CCDF.
}
\label{fig:ccdf-of-questions-and-users}
\end{figure}

\clearpage

\end{document}